\documentclass[aps, prb, twocolumn,amsmath,amssymb,floatfix]{revtex4-2}
\usepackage{graphicx}
\usepackage{physics}
\usepackage{hyperref} 
\hypersetup{colorlinks, allcolors=blue}
\begin{document}

\title{Electron-magnon coupling and quasiparticle lifetimes on the surface of a topological insulator}

\author{Kristian M{\ae}land} 
\affiliation{\mbox{Center for Quantum Spintronics, Department of Physics, Norwegian University of Science and Technology,}\\NO-7491 Trondheim, Norway} 
\author{H{\aa}kon I. R{\o}st} 
\affiliation{\mbox{Center for Quantum Spintronics, Department of Physics, Norwegian University of Science and Technology,}\\NO-7491 Trondheim, Norway} 
\author{Justin W. Wells} 
\affiliation{\mbox{Center for Quantum Spintronics, Department of Physics, Norwegian University of Science and Technology,}\\NO-7491 Trondheim, Norway} 
\author{Asle Sudb{\o}}
\email[Corresponding author: ]{asle.sudbo@ntnu.no}
\affiliation{\mbox{Center for Quantum Spintronics, Department of Physics, Norwegian University of Science and Technology,}\\NO-7491 Trondheim, Norway}

\date{\today} 

\begin{abstract}
The fermionic self-energy on the surface of a topological insulator proximity coupled to ferro- and antiferromagnetic insulators is studied. 
An enhanced electron-magnon coupling is achieved by allowing the electrons on the surface of the topological insulator to have a different exchange coupling to the two sublattices of the antiferromagnet. Such a system is therefore seen as superior to a ferromagnetic interface for the realization of magnon-mediated superconductivity. 
The increased electron-magnon-coupling simultaneously increases the self-energy effects. In this paper we show how the inverse quasiparticle lifetime and energy renormalization on the surface of the topological insulator can be kept low close to the Fermi level by using a magnetic insulator with a sufficient easy-axis anisotropy.
We find that the antiferromagnetic case is most interesting from both a theoretical and an experimental standpoint due to the increased electron-magnon coupling, combined with a reduced need for easy-axis anisotropy compared to the ferromagnetic case.
We also consider a set of material and instrumental parameters where these self-energies should be measurable in angle-resolved photoemission spectroscopy experiments, paving the way for a measurement of the interfacial exchange coupling strength.
\end{abstract}

\maketitle

\section{Introduction}
In conventional Bardeen-Cooper-Schrieffer (BCS) \cite{BCS} superconductors (SC), electron-phonon coupling (EPC) generates an effective, attractive interaction between electrons. Other bosonic excitations also have the capacity to generate attractive electron interactions. For instance, spin fluctuations in magnetic insulators, i.e., magnons, can induce superconductivity in, e.g., normal metals and topological insulators (TI) \cite{KaneMele, KaneHasanTI, QiTITSC2011, he2019topological} by combining materials into heterostructures \cite{EirikTIFMAFM, HenningRex, Hugdal5SCTI, LinderSCTI, kargarianprl, FMNMFMArnePRB, AFMNMAFMArnePRB, thingstad2021eliashberg, EirikSqueezed, EirikSchwinger}. Recently, both BCS- and Amperean-type pairings have been considered as mechanisms for superconductivity on the surface of a TI, exchange coupled to a ferromagnetic insulator (FM) or an antiferromagnetic insulator (AFM) \cite{EirikTIFMAFM, HenningRex}. Such systems have also been studied for other applications, including magnetization dynamics \cite{magetizationdynamics}, confinement of Majorana fermions \cite{FuKane2008, TIFMexp}, magnetoelectric effects \cite{TImagnetoelectric} and proximity induced ferromagnetism \cite{TIFMexp}.

In this paper, we consider the lifetime and energy renormalization of the fermionic quasiparticles on the TI surface in these systems, focusing on the fermion self-energy due to electron-magnon coupling (EMC). Its imaginary part is essentially a measure of the inverse quasiparticle lifetime \cite{BruusFlensberg}, and is used to probe the stability of the fermionic states which underlie the superconducting theories that have been proposed. The real part of the self-energy is used to probe the renormalization of the fermionic states \cite{BruusFlensberg}. A similar study was done in Ref.~\cite{GiraudEgger} for EPC on the surface of an isolated TI. 

In Ref.~\cite{kargarianprl}, the possibility of Amperean pairing was studied for a TI/FM heterostructure, using a self-consistent strong-coupling approach. In the process, the fermion self-energy was studied, and a strong renormalization of the fermionic state was reported. Meanwhile, in Ref.~\cite{EirikTIFMAFM}, superconductivity in both TI/FM and TI/AFM heterostructures were studied within a weak-coupling approach, ignoring any energy renormalizations caused by the magnetic interface \cite{EirikTIFMAFM}. 
In this paper, we reveal for which material parameters the assumption of small renormalization of the fermionic states is permissible. To achieve this, we consider magnetic insulators with an easy-axis anisotropy. 

In the AFM case, we also consider both compensated and uncompensated interfaces. 
Hence, the interfacial exchange coupling to the electrons on the TI surface may be different for the two sublattices of the AFM, where the two sublattices have magnetization ordered in opposite directions \cite{EirikTIFMAFM, AkashOmega}.  
An uncompensated interface, where the electrons on the TI surface couple asymmetrically to the sublattices of the AFM, has been shown to increase EMC, and hence increase the critical temperature for superconductivity \cite{EirikTIFMAFM, EirikSqueezed, EirikSchwinger}. 

On the other hand, a stronger EMC has the potential for more detrimental effects on the fermionic states. We find that the easy-axis anisotropy in the magnetic insulators, and the degree of surface compensation in the AFM case can both be used to increase the lifetime of the fermionic states on the TI surface close to the Fermi level. For sufficient easy-axis anisotropy, we find that it is possible for the fermionic states on the TI surface to remain long-lived and weakly renormalized even when coupled to an uncompensated AFM surface. It is also found that the easy-axis anisotropy needed to stabilize the fermionic states in the AFM case is weaker than that needed in the FM case.

We first consider the case of a TI coupled to a FM in Sec.~\ref{sec:FM}, before moving on to the TI/AFM heterostructure in Sec.~\ref{sec:AFM}. The results for the self-energy and the renormalized Green's function are considered for both systems in Sec.~\ref{sec:Results}. In Sec.~\ref{sec:ARPES}, we examine a new set of material parameters giving self-energies that should be measurable using angle-resolved photoemission spectroscopy (ARPES). The conclusions are given in Sec.~\ref{sec:Con} and the Appendices give further details of the calculations.

\section{Ferromagnet} \label{sec:FM}

\begin{figure}
    \centering
    \includegraphics[width=0.7\linewidth]{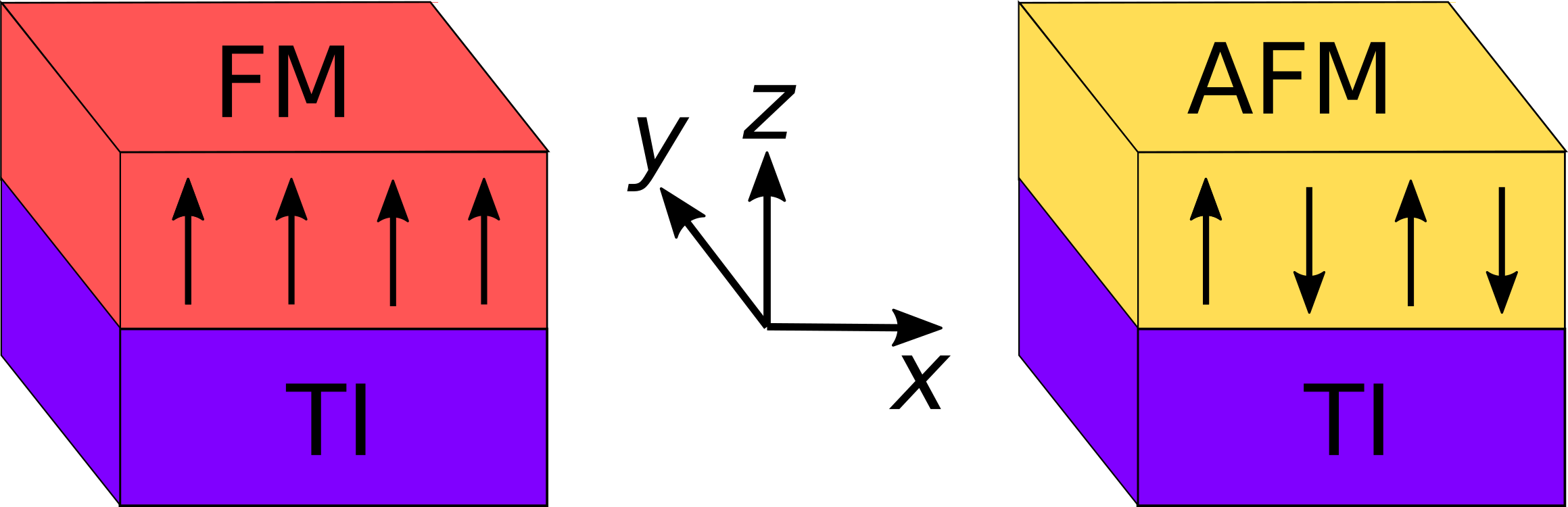}
    \caption{Illustrations of the two heterostructures considered in this paper along with the coordinate system. We consider a topological insulator (TI) coupled to a ferromagnetic insulator (FM) or an antiferromagnetic insulator (AFM).}
    \label{fig:system}
\end{figure}

Our system is a three-dimensional (3D) TI with one surface in contact with a FM, as depicted in the left part of Fig.~\ref{fig:system}. This surface is the $xy$ plane, and we assume an ordered state with magnetization in the $z$ direction in the FM. We consider a TI with one Dirac cone such as $\textrm{Bi}_2\textrm{Se}_3$ or $\textrm{Bi}_2\textrm{Te}_3$ \cite{NatureTIzhang}. Examples of candidate FM materials include yttrium iron garnet (YIG) \cite{ExpFMYIG}, EuO \cite{ExpFMEuO, FMNMFMArnePRB}, and EuS \cite{TIFMexp}. We set $\hbar = k_{\textrm{B}} = a = 1$ in the equations throughout the paper. Here, $\hbar$ is the reduced Planck's constant, $k_{\textrm{B}}$ is Boltzmann's constant, and $a$ is the lattice constant. 

\subsection{Model}
The Hamiltonian describing the interface between the TI and the FM contains a lattice formulation of the TI surface, $H_{\textrm{TI}}$, a Heisenberg model for the FM with an additional easy-axis anisotropy term, $H_{\textrm{FM}}$, and a model for the exchange coupling of lattice site spins in the FM to the electrons on the TI surface, $H_{\textrm{int}}$. We use the same model presented in Ref.~\cite{EirikTIFMAFM}, namely, $H=H_{\textrm{TI}}+H_{\textrm{FM}}+H_{\textrm{int}}$, with
\begin{align}
    H_{\textrm{TI}} =& \frac{v_{\textrm{F}}}{2}\sum_{i} [(\boldsymbol{c}_i^\dagger i\tau_y \boldsymbol{c}_{i+\hat{x}} - \boldsymbol{c}_i^\dagger i\tau_x \boldsymbol{c}_{i+\hat{y}}) + \textrm{H.c.}]  \nonumber \\ 
    &+\sum_{i} \boldsymbol{c}_i^\dagger (2W\tau_z-\mu) \boldsymbol{c}_{i}  \nonumber \\ 
    &-\frac{W}{2} \sum_{i} [(\boldsymbol{c}_i^\dagger \tau_z \boldsymbol{c}_{i+\hat{x}} + \boldsymbol{c}_i^\dagger \tau_z \boldsymbol{c}_{i+\hat{y}}) + \textrm{H.c.}], \label{eq:HTIorig} \\
    H_{\textrm{FM}} =& -J \sum_{\langle i,j \rangle} \boldsymbol{S}_i \cdot \boldsymbol{S}_j - K \sum_{i} S_{iz}^2 , \\
    H_{\textrm{int}} =& -2\Bar{J} \sum_{i} \boldsymbol{c}_i^\dagger \boldsymbol{\tau} \boldsymbol{c}_{i} \cdot \boldsymbol{S}_i.
\end{align}
Here, $v_{\textrm{F}}$ is the Fermi velocity, $\boldsymbol{c}_i^\dagger = (c_{i\uparrow}^\dagger, c_{i\downarrow}^\dagger)$, the fermionic operators $c_{i\sigma}^\dagger$ and $c_{i\sigma}$ create and destroy electrons with spin $\sigma$ at lattice site $i$, respectively, $\boldsymbol{\tau} = (\tau_x, \tau_y, \tau_z)$ are the Pauli matrices, and H.c. denotes the Hermitian conjugate of the preceding term. Furthermore, $\hat{x}$ and $\hat{y}$ are unit vectors in the $x$ direction and $y$ direction, respectively, $\mu$ is the chemical potential, and $\langle i,j \rangle$ indicates that the lattice sites $i$ and $j$ located at $\boldsymbol{r}_i$ and $\boldsymbol{r}_j$ should be nearest neighbors. 
To ease computational requirements, we have assumed a 2D square lattice in the interfacial plane. The first line of $H_{\textrm{TI}}$ represents the spin-momentum locking of electrons on the TI surface. The Wilson terms containing $W$ are added to avoid additional Dirac cones at the Brillouin zone boundaries appearing from a direct discretization of the continuum model \cite{WilsonZhou}. They are added so that the lattice model reproduces the correct physics \cite{EirikTIFMAFM, WilsonZhou}. 
The FM Hamiltonian contains an exchange interaction between nearest-neighbor lattice site spins, $\boldsymbol{S}_i$, with strength $J> 0$, and an easy-axis anisotropy term determined by $K>0$ ensuring that ordering in the $z$ direction is energetically favorable. The interfacial exchange coupling is parametrized by $\Bar{J}$.

The next step is to obtain the fermions which diagonalize the TI Hamiltonian and the magnons which diagonalize the FM Hamiltonian. This was performed in Ref.~\cite{EirikTIFMAFM} and we repeat the main points here. A Holstein-Primakoff (HP) transformation  \cite{HP-PhysRev.58.1098} is introduced for the spin operators $S_{i+} = \sqrt{2S}a_i$, $S_{i-} = \sqrt{2S}a_i^\dagger$, and $S_{iz} = S - a_i^\dagger a_i$, where the bosonic operators $a_i^\dagger$ and $a_i$ create and destroy magnons at lattice site $i$, respectively. 
Furthermore, $S_{i\pm} = S_{ix} \pm i S_{iy}$, while $S$ is the spin quantum number of the lattice site spins. 
Here, we have neglected any terms beyond quadratic in the magnon operators $a_i$, and we continue to do so throughout the analysis. This is permissible when assuming that the spins are nearly ordered, with only small quantum fluctuations, even when $S$ is not large. Additionally, any constant terms in the Hamiltonian are neglected as these merely shift the zero point of the energy. Performing a Fourier transform (FT) on the magnon operators, $a_i = \frac{1}{\sqrt{N}}\sum_{\boldsymbol{q}}a_{\boldsymbol{q}}e^{-i\boldsymbol{q}\cdot \boldsymbol{r}_i}$, where $N$ is the number of lattice sites on the interface, gives
\begin{equation}
    H_{\textrm{FM}} = \sum_{\boldsymbol{q}} \omega_{\boldsymbol{q}} a_{\boldsymbol{q}}^\dagger a_{\boldsymbol{q}},
\end{equation}
with $\omega_{\boldsymbol{q}} = 2KS + 4JS(2-\cos q_x - \cos q_y)$. Notice the gap in the magnon spectrum due to the easy-axis anisotropy, $2KS$. The easy-axis anisotropy stabilizes the ground state with magnetization in the $z$ direction, and a higher energy is needed to excite spin fluctuations. We refer to $\boldsymbol{q}$ as the momentum of the magnon, even though, since we have set $\hbar = a = 1$, it is technically a dimensionless version of the quasimomentum, restricted to the first Brillouin zone (1BZ) of the 2D square lattice.

Inserting the HP transformation, as well as a FT of both the magnon and the electron operators, $c_{i\sigma} = \frac{1}{\sqrt{N}}\sum_{\boldsymbol{k}} c_{\boldsymbol{k}\sigma}e^{-i\boldsymbol{k}\cdot \boldsymbol{r}_i}$, into $H_{\textrm{int}}$ yields
\begin{align}
    H_{\textrm{int}} =& \frac{V}{\sqrt{N}}\sum_{\boldsymbol{k}\boldsymbol{q}} (a_{\boldsymbol{q}}c_{\boldsymbol{k}+\boldsymbol{q}, \downarrow}^\dagger c_{\boldsymbol{k}\uparrow} + a_{-\boldsymbol{q}}^\dagger c_{\boldsymbol{k}+\boldsymbol{q}, \uparrow}^\dagger c_{\boldsymbol{k}\downarrow}) \nonumber \\
    &-2\Bar{J}S\sum_{\boldsymbol{k}\sigma} \sigma c_{\boldsymbol{k}\sigma}^\dagger c_{\boldsymbol{k}\sigma}.
\end{align}
Here, $V = -2\Bar{J}\sqrt{2S}$, $\sigma = 1$ for spin up, and $\sigma = -1$ for spin down. The first line describes EMC involving one magnon, while terms showing EMC with more than one magnon have been neglected. We have also neglected any Umklapp processes, since a small Fermi surface close to the 1BZ center means the Fermi momentum is much smaller than the reciprocal lattice vectors \cite{EirikTIFMAFM}.
The terms in the second line do not contain magnon operators and are therefore moved to $H_{\textrm{TI}}$. 
These terms act like an external magnetic field on the TI and will open a gap in the fermionic spectrum. This is due to the magnetization along the $z$ direction in the FM.

\begin{figure}
    \centering
    \includegraphics[width=0.9\linewidth]{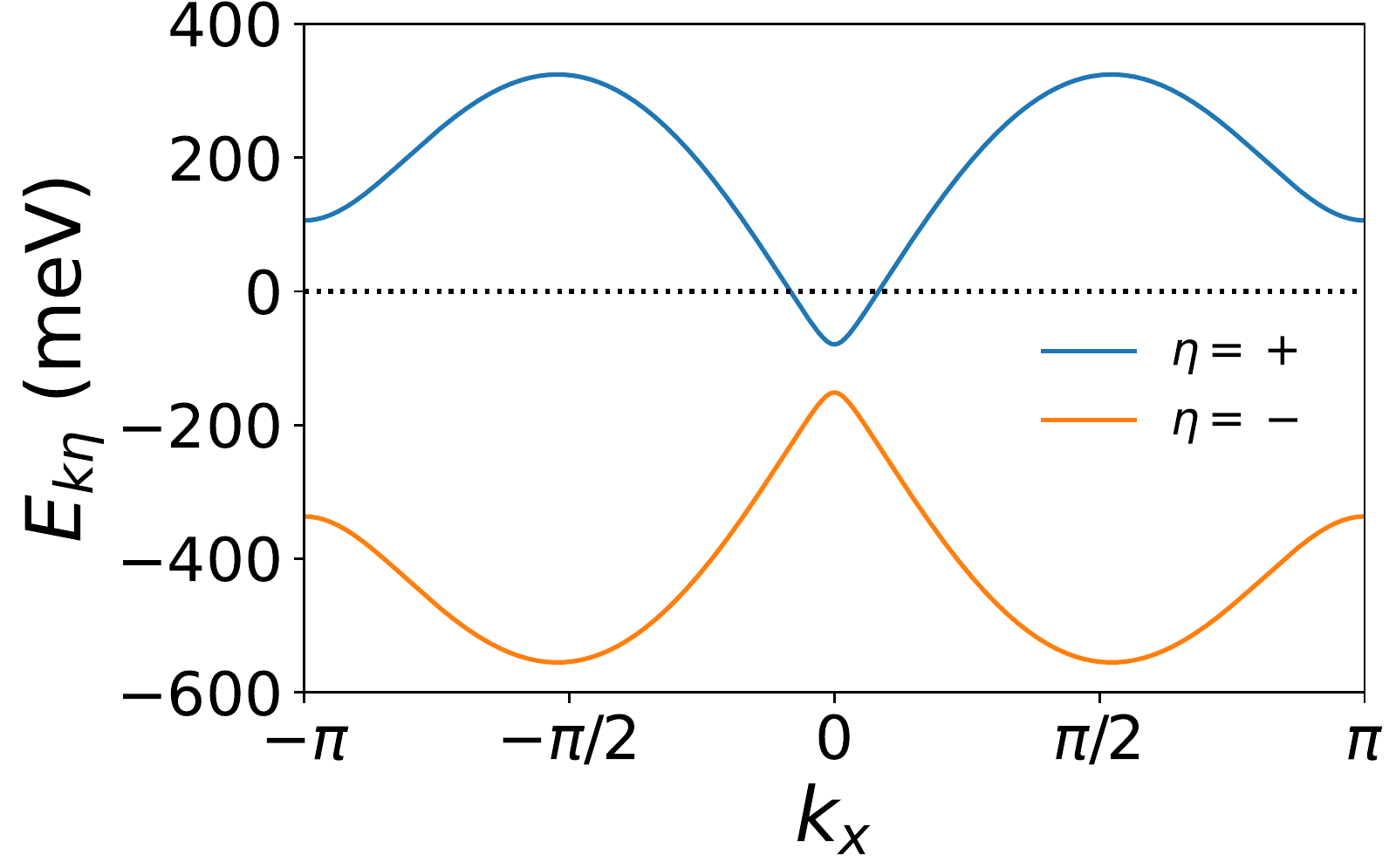}
    \caption{The TI excitation spectrum $E_{\boldsymbol{k}\eta}$ for the TI/FM heterostructure plotted along $k_x$ with $k_y = 0$, $k_{\textrm{F}} = \pi/12$, $v_{\textrm{F}} = 429~$meV, $W=0.3v_{\textrm{F}}$, $\Bar{J} = 18~$meV, and $S=1$. The chemical potential is $\mu \approx 115~$meV.}
    \label{fig:Ek}
\end{figure}

Next, the electron operators in $H_{\textrm{TI}}$ are FT and a unitary transformation is used to diagonalize the Hamiltonian in terms of quasiparticles $\psi_{\boldsymbol{k}\eta}$ with the helicity band index $\eta = \pm$,
\begin{equation}
     H_{\textrm{TI}} = \sum_{\boldsymbol{k} \eta} E_{\boldsymbol{k}\eta} \psi_{\boldsymbol{k}\eta}^\dagger \psi_{\boldsymbol{k}\eta}.
\end{equation}
The excitation energies are $E_{\boldsymbol{k}\eta} = -\mu + \eta F_{\boldsymbol{k}}$, where we define $B_{\boldsymbol{k}} = W(2-\cos k_x - \cos k_y) -2\Bar{J}S$, $C_{\boldsymbol{k}} = -v_{\textrm{F}} \sin k_y$, $D_{\boldsymbol{k}} = -v_{\textrm{F}} \sin k_x$, and $F_{\boldsymbol{k}} = \sqrt{B_{\boldsymbol{k}}^2 + C_{\boldsymbol{k}}^2 + D_{\boldsymbol{k}}^2}$.
Also defining $N_{\boldsymbol{k}} = 2F_{\boldsymbol{k}}(F_{\boldsymbol{k}} +B_{\boldsymbol{k}})$, the electron operators are related to the quasiparticle operators as
\begin{align}
    c_{\boldsymbol{k}\uparrow} &= Q_{\uparrow +}(\boldsymbol{k}) \psi_{\boldsymbol{k}+} + Q_{\uparrow -}(\boldsymbol{k})\psi_{\boldsymbol{k}-}, \\
    c_{\boldsymbol{k}\downarrow} &= Q_{\downarrow +}(\boldsymbol{k}) \psi_{\boldsymbol{k}+} + Q_{\downarrow -}(\boldsymbol{k})\psi_{\boldsymbol{k}-},
\end{align}
with transformation coefficients
\begin{align}
\label{eq:Qcoeff}
    Q_{\uparrow +}(\boldsymbol{k}) &= -Q_{\downarrow -}(\boldsymbol{k}) = (F_{\boldsymbol{k}} +B_{\boldsymbol{k}})/\sqrt{N_{\boldsymbol{k}}}, \\
    Q_{\uparrow -}(\boldsymbol{k}) &= \phantom{-}Q_{\downarrow +}^*(\boldsymbol{k}) = (C_{\boldsymbol{k}} +iD_{\boldsymbol{k}})/\sqrt{N_{\boldsymbol{k}}}. \label{eq:Qcoeff2}
\end{align}
The TI excitation spectrum is plotted in Fig.~\ref{fig:Ek}. The exchange coupling has introduced a gap of $4\Bar{J}S$ in the original Dirac cone, similar to a mass gap for massive Dirac fermions \cite{DiracMaterials}. The Wilson terms open gaps at the boundaries of the 1BZ, ensuring that there is only one Dirac cone present in the system \cite{WilsonZhou}.

Finally transforming $H_{\textrm{int}}$ to the basis which diagonalizes $H_{\textrm{TI}}$ gives
\begin{align}
\label{eq:HintFM}
        H_{\textrm{int}} =& \frac{V}{\sqrt{N}} \sum_{\boldsymbol{k} \boldsymbol{q}}\sum_{\eta\eta'} \big[  Q_{\downarrow \eta}^* (\boldsymbol{k}+\boldsymbol{q}) Q_{\uparrow \eta'}(\boldsymbol{k})  a_{\boldsymbol{q}} \psi_{\boldsymbol{k}+\boldsymbol{q}, \eta}^\dagger \psi_{\boldsymbol{k}\eta'}  \nonumber \\
        &+ Q_{\uparrow \eta}^* (\boldsymbol{k}+\boldsymbol{q}) Q_{\downarrow \eta'}(\boldsymbol{k})  a_{-\boldsymbol{q}}^\dagger\psi_{\boldsymbol{k}+\boldsymbol{q}, \eta}^\dagger \psi_{\boldsymbol{k}\eta'} \big].
\end{align}

\subsection{Self-energy}
\begin{figure}
    \centering
    \begin{minipage}{.48\textwidth}
      \includegraphics[width=\linewidth]{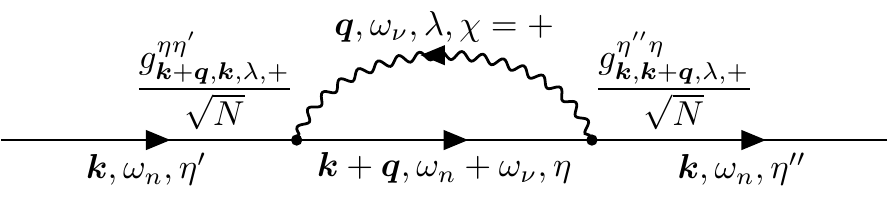}
      \label{fig:Feynman+}
    \end{minipage}%
    \\
    \begin{minipage}{.48\textwidth}
      \includegraphics[width=\linewidth]{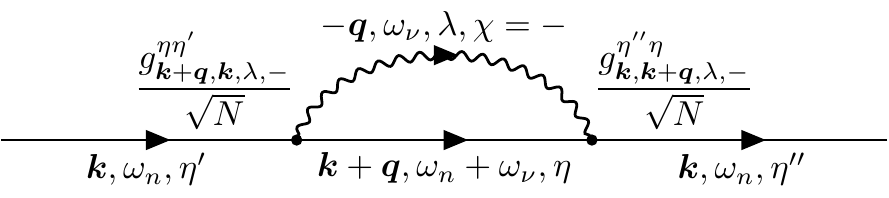}
      \label{fig:feynman-}
    \end{minipage}%
    \caption{The sunset Feynman diagrams considered in this paper. The straight lines represent fermions, while the wavy lines are magnons. We refer to the text for explanations of the symbols. The external lines are included since the incoming and outgoing fermions influence the coupling constants, but their propagators are not included in the self-energy.   \label{fig:Feynman}}
\end{figure}

At this point our calculations diverge from those of Ref.~\cite{EirikTIFMAFM}, as we now calculate the self-energy of the fermionic quasiparticles due to EMC.
We include nonzero temperature by going to a sum over Matsubara frequencies and find \cite{BruusFlensberg, abrikosov, mahanMatsubara, GiraudEgger, elphcselfenergyprb, elphcselfenergyprlpark, elphcselfenergyprltse}
\begin{align}
\label{eq:sigma}
    \Sigma^{\eta''\eta'}(\boldsymbol{k}&, i\omega_n) = - \sum_{\boldsymbol{q}}\sum_{\eta, \lambda, \chi = \pm} \frac{g_{\boldsymbol{k}+\boldsymbol{q}, \boldsymbol{k}, \lambda, \chi}^{\eta\eta'} g_{\boldsymbol{k}, \boldsymbol{k}+\boldsymbol{q}, \lambda, \chi}^{\eta''\eta}}{N} \nonumber \\ 
    &\cross T\sum_{\omega_\nu} D_0^\chi(\boldsymbol{q},i\omega_\nu)G_0^\eta(\boldsymbol{k}+\boldsymbol{q}, i\omega_n+i\omega_\nu),
\end{align}
based on the sunset Feynman diagrams presented in Fig.~\ref{fig:Feynman}. Hence, we have truncated our calculation of the self-energy at second order in the EMC, employing the Migdal approximation \cite{migdal, elphcselfenergyprlpark}. We have also used the fact that the tadpole diagram gives zero contribution in the systems considered in this paper, as shown in Appendix~\ref{sec:tadpole}. $\chi$ labels the direction of the magnon, i.e., the sign in front of $\boldsymbol{q}$, 
while $\lambda$ labels the magnon mode, which for the FM case is superfluous. Based on Eq.~\eqref{eq:HintFM} we have, e.g., the coupling constant $g_{\boldsymbol{k}+\boldsymbol{q}, \boldsymbol{k}, \chi = +}^{\eta\eta'} = V Q_{\downarrow \eta}^*(\boldsymbol{k}+\boldsymbol{q}) Q_{\uparrow \eta'}(\boldsymbol{k})$.

The bare fermion Green's function is \cite{abrikosov, BruusFlensberg} $G_0^{\eta}(\boldsymbol{k}, i\omega_n) = 1/(i\omega_n-E_{\boldsymbol{k}\eta})$, with $\omega_n = (2n+1)\pi T$. Upon introducing $\omega_\nu = 2\pi \nu T$, we use the bare magnon Green's function \cite{BruusFlensberg} 
$D_0^{\chi} (\boldsymbol{q}, i\omega_\nu) = \chi/(i\omega_\nu -\chi \omega_{\boldsymbol{q}})$
for the magnon operator $a_{\boldsymbol{q}}$ when $\chi = +$ and $a_{-\boldsymbol{q}}^\dagger$ when $\chi = -$, both with dispersion $\omega_{\boldsymbol{q}}$. Hence, we use separate propagators for $a_{\boldsymbol{q}}$ and $a_{-\boldsymbol{q}}^\dagger$ as opposed to, e.g., EPC where one usually finds the propagator of the sum of these \cite{elphcselfenergyprltse}. 
In other words, the magnons moving forward and backward in time are treated separately, before adding their respective contributions.

We first perform the sum over the Matsubara frequencies \cite{BruusFlensberg, mahanMatsubara},
\begin{align}
    -&T\sum_{\omega_\nu} \frac{\chi}{i\omega_\nu -\chi \omega_{\boldsymbol{q}}}\frac{1}{i\omega_\nu+i\omega_n-E_{\boldsymbol{k}+\boldsymbol{q}, \eta}}  \nonumber\\
    &=\frac{\chi}{i\omega_n - E_{\boldsymbol{k}+\boldsymbol{q}, \eta} +\chi \omega_{\boldsymbol{q}}} \left[ B_{\text{E}}(\chi \omega_{\boldsymbol{q}}) + F_{\text{D}}(E_{\boldsymbol{k}+\boldsymbol{q}, \eta})\right].
\end{align}
Here, $B_{\text{E}}(\epsilon) = 1/(e^{\epsilon/T}-1) = [\coth(\epsilon /2T)-1]/2$ is the Bose-Einstein distribution and $F_{\text{D}}(\epsilon) = 1/(e^{\epsilon/T}+1) = [1-\tanh(\epsilon /2T)]/2$ is the Fermi-Dirac distribution. 
Using $B_{\text{E}}(-\epsilon) = -1 - B_{\text{E}}(\epsilon)$ and an analytic continuation $i\omega_n \to \omega + i\delta$, where $\delta = 0^+$ \cite{mahanMatsubara, elphcselfenergyprlpark, elphcselfenergyprltse}, yields
\begin{align}
\label{eq:Matsubarasum}
        -&T\sum_{\omega_\nu}D_0^{\chi}(\boldsymbol{q},i\omega_\nu)G_0^\eta(\boldsymbol{k}+\boldsymbol{q}, i\omega_n+i\omega_\nu)  \nonumber \\
        &=\frac12 \frac{\coth\frac{\omega_{\boldsymbol{q}}}{2T}-\chi\tanh\frac{E_{\boldsymbol{k}+\boldsymbol{q},\eta}}{2T}}{\omega-E_{\boldsymbol{k}+\boldsymbol{q}, \eta}+\chi\omega_{\boldsymbol{q}}+i\delta}. 
\end{align}
The following transformation is used in the sum over momentum:
\begin{equation}
\label{eq:FMtransform}
    \sum_{\boldsymbol{q}} \to \frac{N}{(2\pi)^2}\int_{-\pi}^{\pi} dq_x \int_{-\pi}^{\pi} dq_y. 
\end{equation}
Inserting Eqs.~\eqref{eq:Matsubarasum} and \eqref{eq:FMtransform} into Eq.~\eqref{eq:sigma} gives 
\begin{align}
\label{eq:sigmaFM}
     \Sigma&^{\eta''\eta'}(\boldsymbol{k}, \omega) = \sum_{\eta, \chi = \pm} \frac{1}{8\pi^2}\int_{-\pi}^{\pi} dq_x \int_{-\pi}^{\pi} dq_y \nonumber \\
     &\cross g_{\boldsymbol{k}+\boldsymbol{q}, \boldsymbol{k}, \chi}^{\eta\eta'} g_{\boldsymbol{k}, \boldsymbol{k}+\boldsymbol{q}, \chi}^{\eta''\eta} \frac{\coth\frac{\omega_{\boldsymbol{q}}}{2T}-\chi\tanh\frac{E_{\boldsymbol{k}+\boldsymbol{q},\eta}}{2T}}{\omega-E_{\boldsymbol{k}+\boldsymbol{q}, \eta}+\chi\omega_{\boldsymbol{q}}+i\delta}.
\end{align}

\subsubsection{Imaginary part of the self-energy}
In the following, we focus on the case where $\eta'' = \eta'$. Then, $ g_{\boldsymbol{k}, \boldsymbol{k}+\boldsymbol{q}, \chi}^{\eta''\eta} = (g_{\boldsymbol{k}+\boldsymbol{q}, \boldsymbol{k}, \chi}^{\eta\eta'})^*$ and the coupling constant factor is real.
Hence,
\begin{align}
\label{eq:imsigmaFM}
    &\Im\Sigma^{\eta'\eta'}(\boldsymbol{k}, \omega) =  \sum_{\eta, \chi = \pm} \frac{-1}{8\pi}\int_{-\pi}^{\pi} dq_x \int_{-\pi}^{\pi} dq_y \abs{g_{\boldsymbol{k}+\boldsymbol{q}, \boldsymbol{k}, \chi}^{\eta\eta'}}^2 \nonumber \\
    & \cross \delta(\omega-E_{\boldsymbol{k}+\boldsymbol{q},\eta}+\chi\omega_{\boldsymbol{q}}) \left(\coth\frac{\omega_{\boldsymbol{q}}}{2T}-\chi\tanh\frac{E_{\boldsymbol{k}+\boldsymbol{q},\eta}}{2T}\right).
\end{align}
Transforming to polar coordinates yields
\begin{align}
    &\Im\Sigma^{\eta'\eta'}(\boldsymbol{k}, \omega) = \sum_{\eta, \chi = \pm}\frac{-1}{8\pi}\int_{-\pi}^\pi d\theta \int_0^{c(\theta)} dq q \abs{g_{\boldsymbol{k}, q, \theta, \chi}^{\eta\eta'}}^2  \nonumber\\
    &\cross \delta(\omega-E_{\boldsymbol{k},q, \theta, \eta}+\chi \omega_{q,\theta}) \left(\coth\frac{\omega_{q,\theta}}{2T}-\chi \tanh\frac{E_{\boldsymbol{k},q, \theta, \eta}}{2T}\right).
\end{align}
Here, $q_x = q\cos\theta$ and $q_y = q\sin\theta$. The upper cutoff $c(\theta) = \pi/\operatorname{max}(|\sin\theta|,|\cos\theta|)$ ensures that the integral is limited to the 1BZ.
We calculate this integral using \cite{integraldelta}
\begin{equation}
    \delta(f(r)) = \sum_i \frac{\delta(r-r_i)}{\abs{f'(r_i)}},
\end{equation}
for a continuously differentiable function $f(r)$ with roots $r_i$ and where $f'(r_i) \neq 0$. Here, the expression inside the $\delta$ function is $f_{\eta\chi}(q) = \omega-E_{\boldsymbol{k},q, \theta, \eta}+\chi \omega_{q,\theta}$. Its roots are found numerically, labeled $q_i$ if they satisfy $ 0 \leq q_i \leq c(\theta)$, and ignored otherwise.
Integrating over $q$ gives
\begin{align}
\label{eq:numericimsigma}
    \Im\Sigma^{\eta'\eta'}(\boldsymbol{k}&, \omega) = \sum_{\eta, \chi = \pm}\frac{-1}{8\pi}\int_{-\pi}^\pi d\theta  \sum_{i} \frac{\abs{g_{\boldsymbol{k}, q_i, \theta, \chi}^{\eta\eta'}}^2 }{\abs{f_{\eta\chi}'(q_i)}} \nonumber \\
    &\cross q_i\left(\coth\frac{\omega_{q_i,\theta}}{2T}-\chi \tanh\frac{E_{\boldsymbol{k},q_i, \theta, \eta}}{2T}\right).
\end{align}
Some details of this treatment of the $\delta$ function are commented on in Appendix~\ref{sec:delta}.

\subsubsection{Real part of the self-energy}
The real part of the self-energy can be found using the Kramers-Kronig relations \cite{Kramers-Kronig},
\begin{equation}
\label{eq:resigma}
    \Re\Sigma^{\eta'\eta'}(\boldsymbol{k}, \omega) = \frac{1}{\pi}P\int_{-\infty}^{\infty}\frac{\Im\Sigma^{\eta'\eta'}(\boldsymbol{k}, \omega')}{\omega'-\omega}d\omega',
\end{equation}
with $P$ indicating the Cauchy principal value.
Here, this integral is calculated using the trapezoidal rule. An important consideration is that the points that are chosen for $\omega'$ are evenly distributed around the singularity at $\omega$.

\section{Antiferromagnet} \label{sec:AFM}

We now replace the FM with an AFM and assume a staggered state with magnetization along the $z$ direction on the bipartite lattice of the AFM. The system is illustrated in the right part of Fig.~\ref{fig:system}. Examples of candidate AFM materials include $\textrm{Cr}_2\textrm{O}_3$ \cite{ExpAFMCr2O3}, $\textrm{Fe}_2\textrm{O}_3$ \cite{ExpAFMFe2O3}, and $\textrm{MnF}_2$ \cite{ExpAFMMnF2, AFMNMAFMArnePRB}.   

\subsection{Model}
\begin{figure}
    \centering
    \includegraphics[width=0.7\linewidth]{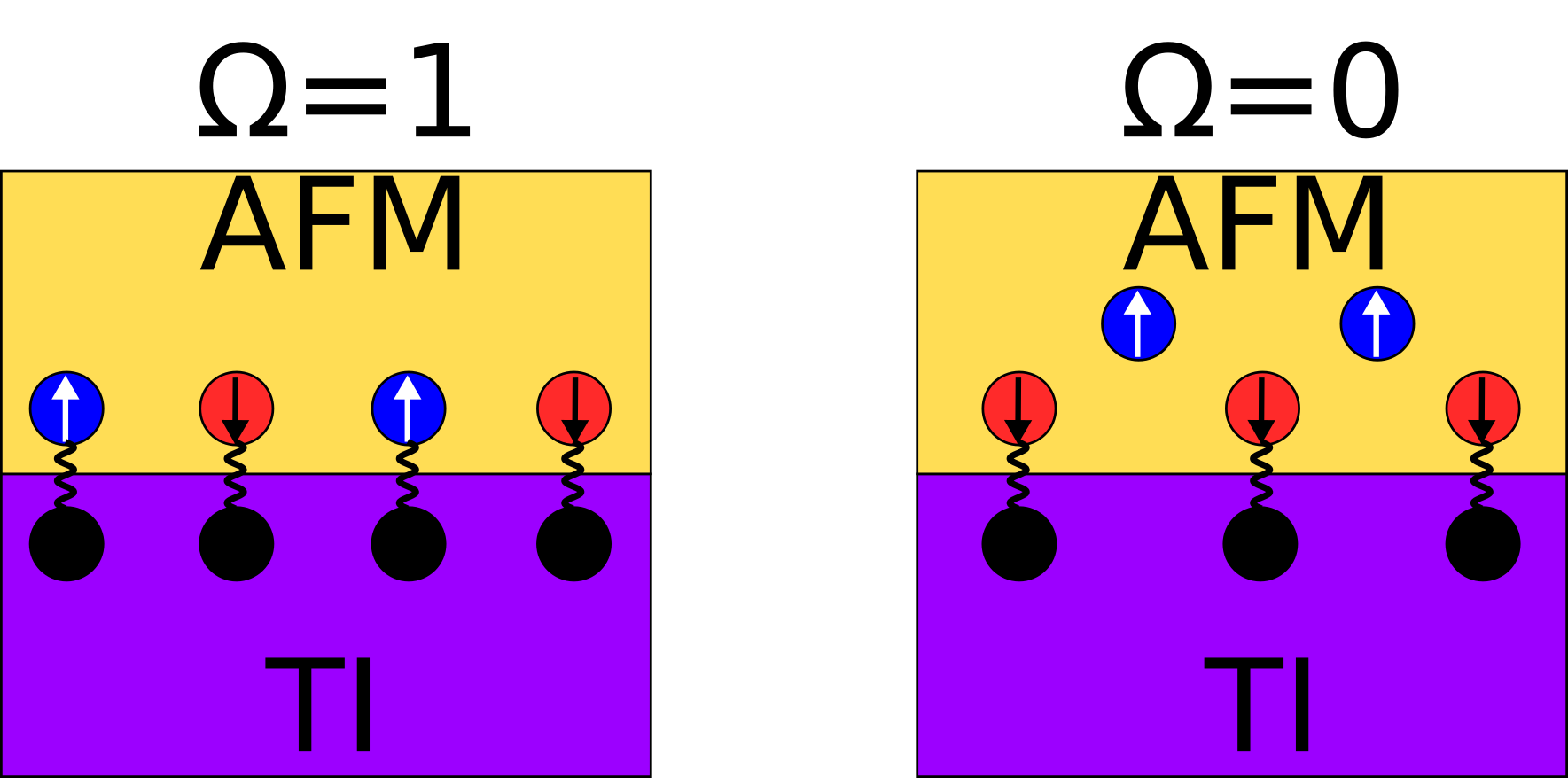}
    \caption{An illustration of the interfacial exchange coupling between electrons in the TI and lattice site spins in the AFM. For a compensated interface, with $\Omega \equiv \Bar{J}_A/\Bar{J}_B = 1$, we have an equal coupling to both sublattices. For a completely uncompensated interface, with $\Omega = 0$, the electrons in the TI couple to only one of the sublattices in the AFM. The figure is inspired by Ref.~\cite{EirikTIFMAFM}.}
    \label{fig:interfaceOmega}
\end{figure}

We use the same model presented in Ref.~\cite{EirikTIFMAFM}, namely, $H = H_{\textrm{TI}} + H_{\textrm{AFM}}+ H_{\textrm{int}}$, with
\begin{align}
    H_{\textrm{AFM}} =& J_1 \sum_{\langle i,j \rangle} \boldsymbol{S}_i \cdot \boldsymbol{S}_j + J_2 \sum_{\langle\langle i,j \rangle\rangle} \boldsymbol{S}_i \cdot \boldsymbol{S}_j - K \sum_{i} S_{iz}^2 , \\
    H_{\textrm{int}} =& -2\Bar{J}_A \sum_{i\in A} \boldsymbol{c}_i^\dagger \boldsymbol{\tau} \boldsymbol{c}_{i} \cdot \boldsymbol{S}_i-2\Bar{J}_B \sum_{i\in B} \boldsymbol{c}_i^\dagger \boldsymbol{\tau} \boldsymbol{c}_{i} \cdot \boldsymbol{S}_i,
\end{align}
and $H_{\textrm{TI}}$ as in Eq.~\eqref{eq:HTIorig}. Here,  $\langle\langle i,j \rangle\rangle$ indicates  that the lattice sites $i$ and $j$ should be next-nearest neighbors. Once again, we have assumed a 2D square lattice in the interfacial plane for computational convenience. The AFM Hamiltonian contains an exchange interaction between nearest-neighbor lattice site spins with strength $J_1> 0$ and between next-nearest neighbors with strength $J_2$. If $J_2 < 0$ this term stabilizes the AFM state, while if $J_2>0$ it acts as a frustration. We assume $|J_2| \ll J_1$ such that the system remains in the staggered state also for $J_2>0$. The easy-axis anisotropy term is the same as in the FM case. The sublattices of the bipartite lattice in the AFM are labeled $A$ and $B$. The exchange coupling to the electrons on the TI surface is parametrized by $\Bar{J}_A$ and $\Bar{J}_B$ for lattice site spins on the $A$ and $B$ sublattices, respectively. We allow $\Bar{J}_A$ and $\Bar{J}_B$ to be different, which can describe an uncompensated antiferromagnetic interface where one sublattice is more exposed than the other \cite{EirikTIFMAFM}. This is illustrated in Fig.~\ref{fig:interfaceOmega}. 
We introduce $\Bar{J} \equiv \Bar{J}_B$ and $\Omega \equiv \Bar{J}_A/\Bar{J}_B$, and let $0 \leq \Omega \leq 1$ parametrize the sublattice asymmetry of the exchange coupling.

Obtaining the eigenexcitations of the TI and the AFM follows a similar methodology as the FM case \cite{EirikTIFMAFM}, and we focus on the main differences. We assume the lattice site spins on the $A$ sublattice point in the positive $z$ direction and opposite alignment on the $B$ sublattice. A HP transformation is introduced for the spin operators $S_{i+}^A = \sqrt{2S}a_i$, $S_{i-}^A = \sqrt{2S}a_i^\dagger$, $S_{iz}^A = S - a_i^\dagger a_i$, $S_{i+}^B = \sqrt{2S}b_i^\dagger$, $S_{i-}^B = \sqrt{2S}b_i$, and $S_{iz}^B = -S + b_i^\dagger b_i$. Next, we introduce FT of the magnon operators, $a_i = \frac{1}{\sqrt{N_A}}\sum_{\boldsymbol{q}\in\Diamond}a_{\boldsymbol{q}}e^{-i\boldsymbol{q}\cdot \boldsymbol{r}_i}$ and $b_i = \frac{1}{\sqrt{N_B}}\sum_{\boldsymbol{q}\in\Diamond}b_{\boldsymbol{q}}e^{-i\boldsymbol{q}\cdot \boldsymbol{r}_i}$. Here, $N_A$ and $N_B$ are the number of lattice sites in the sublattices, and we assume $N_A = N_B = N/2$, where $N$ is the total number of lattice sites on the interface. The sums over $\boldsymbol{q}$ are restricted to the reduced Brillouin zone (RBZ) of the sublattices, which is indicated by $\boldsymbol{q} \in \Diamond$. $H_{\textrm{AFM}}$ is not diagonal in the original sublattice magnons $a_{\boldsymbol{q}} $ and $b_{\boldsymbol{q}} $. Hence, a Bogoliubov transformation is introduced, expressing new magnon operators as $\alpha_{\boldsymbol{q}+} = u_{\boldsymbol{q}}a_{\boldsymbol{q}} - v_{\boldsymbol{q}}b_{-\boldsymbol{q}}^\dagger $ and $\alpha_{\boldsymbol{q}-} =  u_{\boldsymbol{q}}b_{\boldsymbol{q}} - v_{\boldsymbol{q}}a_{-\boldsymbol{q}}^\dagger$. Requiring that the new operators are bosonic fixes $|u_{\boldsymbol{q}}|^2-|v_{\boldsymbol{q}}|^2 = 1$. We assume $u_{\boldsymbol{q}}$ and $v_{\boldsymbol{q}}$ are real, as well as inversion symmetric in $\boldsymbol{q}$. Requiring that the AFM Hamiltonian is diagonal in terms of these new magnon operators yields
\begin{equation}
    H_{\textrm{AFM}} = \sum_{\boldsymbol{q}\in\Diamond} \sum_{\lambda = \pm} \omega_{\boldsymbol{q}} \alpha_{\boldsymbol{q}\lambda}^\dagger\alpha_{\boldsymbol{q}\lambda},
\end{equation}
with $\omega_{\boldsymbol{q}} = \sqrt{\lambda_{\boldsymbol{q}}^2-\gamma_{\boldsymbol{q}}^2}$, $\lambda_{\boldsymbol{q}} = 2KS + 8J_1S + 8J_2S(\cos q_x \cos q_y-1)$, and $\gamma_{\boldsymbol{q}} = 4J_1 S (\cos q_x + \cos q_y)$. The gap in the AFM magnon spectrum, $\omega_{\boldsymbol{q}=0} = \sqrt{32J_1KS^2+4K^2S^2}$, is significantly greater than the gap in the FM case, $2KS$, provided $K\ll J \approx J_1$.
On the other hand, even with comparable gaps, there are far more low-energy magnons in the FM than in the AFM. The reason is that the ungapped FM spectrum is quadratic for small $|\boldsymbol{q}|$, while the ungapped AFM spectrum is linear for small $|\boldsymbol{q}|$.

The FT of the electron operators is now written $c_{i\sigma} = \frac{1}{\sqrt{N}}\sum_{\boldsymbol{k}\in \square} c_{\boldsymbol{k}\sigma}e^{-i\boldsymbol{k}\cdot \boldsymbol{r}_i}$, where $\boldsymbol{k}\in \square$ indicates that the sum runs over the entire 1BZ. Inserting the HP transformation, as well as a FT of both the magnon and the electron operators, into $H_{\textrm{int}}$ yields some terms describing EMC and some terms that do not contain magnon operators. The latter terms are included in $H_{\textrm{TI}}$, similar to the FM case.

Next, the electron operators in $H_{\textrm{TI}}$ are FT and a unitary transformation is used to diagonalize the Hamiltonian in terms of quasiparticles $\psi_{\boldsymbol{k}\eta}$ with helicity index $\eta = \pm$,
\begin{equation}
     H_{\textrm{TI}} = \sum_{\boldsymbol{k}\in \square, \eta} E_{\boldsymbol{k}\eta} \psi_{\boldsymbol{k}\eta}^\dagger \psi_{\boldsymbol{k}\eta}.
\end{equation}
The definition of $B_{\boldsymbol{k}}$ is changed to $B_{\boldsymbol{k}} = W(2-\cos k_x - \cos k_y) -\Bar{J}S(\Omega-1)$, while the other definitions remain the same as in the FM case. As opposed to the FM case, this means that the fermion spectrum can be ungapped if $\Omega = 1$, since, with equal coupling to the two sublattices, no net magnetization affects the TI. Additionally, the fermion gap is smaller for the same $\Bar{J}$ with $\Omega = 0$, since a smaller net magnetization affects the TI surface.

Finally transforming $H_{\textrm{int}}$ to the bases which diagonalize $H_{\textrm{TI}}$ and $H_{\textrm{AFM}}$ gives
\begin{align}
\label{eq:HintAFM}
    H_{\textrm{int}} = \frac{U}{\sqrt{N}} \sum_{\begin{smallmatrix} \boldsymbol{k} \in \square \\  \boldsymbol{q} \in \Diamond \\ \eta\eta'  \end{smallmatrix}} \Big\{ &\big[ (\Omega u_{\boldsymbol{q}} + v_{\boldsymbol{q}}) \alpha_{\boldsymbol{q}+} + (\Omega v_{\boldsymbol{q}}+u_{\boldsymbol{q}})\alpha_{-\boldsymbol{q},-}^\dagger \big] \nonumber \\
    &\cross Q_{\downarrow \eta}^* (\boldsymbol{k}+\boldsymbol{q}) Q_{\uparrow \eta'}(\boldsymbol{k})  \psi_{\boldsymbol{k}+\boldsymbol{q}, \eta}^\dagger \psi_{\boldsymbol{k}\eta'}  \nonumber \\
    +&\big[ (\Omega u_{\boldsymbol{q}} + v_{\boldsymbol{q}}) \alpha_{-\boldsymbol{q},+}^\dagger + (\Omega v_{\boldsymbol{q}}+u_{\boldsymbol{q}})\alpha_{\boldsymbol{q}-} \big] \nonumber \\
    &\cross Q_{\uparrow \eta}^* (\boldsymbol{k}+\boldsymbol{q}) Q_{\downarrow \eta'}(\boldsymbol{k})  \psi_{\boldsymbol{k}+\boldsymbol{q}, \eta}^\dagger \psi_{\boldsymbol{k}\eta'} \Big\},
\end{align}
where $U = -2\Bar{J}\sqrt{S}$.


The factors in the Bogoliubov transformation are 
\begin{align}
\label{eq:Bogoliubov_u}
    u_{\boldsymbol{q}} &= \sqrt{\lambda_{\boldsymbol{q}}/2\omega_{\boldsymbol{q}}+1/2}, \\
    v_{\boldsymbol{q}} &= \operatorname{sgn}(-\gamma_{\boldsymbol{q}}/\lambda_{\boldsymbol{q}})\sqrt{u_{\boldsymbol{q}}^2-1}. \label{eq:Bogoliubov_v}
\end{align}
As it turns out, $v_{\boldsymbol{q}} \approx -u_{\boldsymbol{q}}$ when $\boldsymbol{q} \to 0$, an approximation which becomes better as $K \to 0$. Hence, the combinations like $\Omega u_{\boldsymbol{q}} + v_{\boldsymbol{q}}$ appearing in the EMC Hamiltonian in Eq.~\eqref{eq:HintAFM} are small for $\Omega = 1$, i.e., a compensated AFM interface with equal coupling to both sublattices, while they can be very large for $\Omega = 0$, i.e., a totally uncompensated AFM interface where the electrons on the TI surface couple to only one sublattice. This was also explained in Ref.~\cite{EirikTIFMAFM}, where Fig.~7, in addition to plotting $u_{\boldsymbol{q}}$ and $v_{\boldsymbol{q}}$, shows how a positive $J_2$, i.e., a frustration of the AFM, can also increase the coupling. We therefore consider $J_2 = 0.05J_1$ in this paper. 

Another point is that if the easy-axis anisotropy parameter $K$ is removed, $\lim_{\boldsymbol{q}\to 0}u_{\boldsymbol{q}} = \infty$ and $\lim_{\boldsymbol{q}\to 0}v_{\boldsymbol{q}} = -\infty$. Hence, the coupling constants of the EMC would be infinite at $\boldsymbol{q} = 0$ if $\Omega \neq 1$. This in turn would lead to a divergent self-energy within the presented framework. This divergent behavior might be removed by using a self-consistent approach where renormalized propagators are used in calculating the self-energy. It may also be necessary to include higher-order diagrams in the calculations. We will continue to use the bare propagators and truncate at second order in EMC. Therefore, we will keep $K > 0$, introducing a gap in the magnon spectrum as well as making $u_{\boldsymbol{q}=0}$ and $v_{\boldsymbol{q}=0}$ finite. A similar divergence would also occur for the TI/FM heterostructure at $K=0$ within the presented framework. There, the reason is that $\lim_{q\to0} q\coth(\omega_{q, \theta}/2T) = \infty$ since $\omega_{q, \theta}$ is quadratic for small $q$ and ungapped when $K=0$.

\subsection{Self-energy}
With two magnon modes due to the presence of two sublattices, we now keep the sum over $\lambda = \pm$ in Eq.~\eqref{eq:sigma}. Based on Eq.~\eqref{eq:HintAFM} we have, e.g., $g_{\boldsymbol{k}+\boldsymbol{q}, \boldsymbol{k}, \lambda = +, \chi = +}^{\eta\eta'} = U(\Omega u_{\boldsymbol{q}}+ v_{\boldsymbol{q}})Q_{\downarrow \eta}^*(\boldsymbol{k}+\boldsymbol{q}) Q_{\uparrow \eta'}(\boldsymbol{k})$. The expressions for the magnon propagators are unchanged, apart from a redefinition of the magnon spectrum, and are now applied to the magnon operators $\alpha_{\boldsymbol{q}\lambda}$ when $\chi = +$ and $\alpha_{-\boldsymbol{q},\lambda}^\dagger$ when $\chi = -$, all with dispersion $\omega_{\boldsymbol{q}}$. The sum over momentum is transformed as
\begin{align}
    \sum_{\boldsymbol{q} \in \Diamond} &\to \frac{N}{(2\pi)^2}\int_{-\pi}^{\pi} dq_x \int_{-\pi+|q_x|}^{\pi-|q_x|} dq_y \nonumber \\
    &\to \frac{N}{(2\pi)^2} \int_{-\pi}^{\pi} d\theta \int_0^{c(\theta)} dq q.
\end{align}
Here, $q_x = q\cos\theta$ and $q_y = q\sin\theta$. The upper cutoff $c(\theta) = \pi/(|\sin\theta|+|\cos\theta|)$ ensures that the integral is limited to the RBZ.

Otherwise proceeding just as in the FM case gives
\begin{align}
\label{eq:numericimsigmaAFM}
    \Im\Sigma^{\eta'\eta'}(\boldsymbol{k}&, \omega) = \sum_{\eta, \lambda, \chi = \pm}\frac{-1}{8\pi}\int_{-\pi}^\pi d\theta  \sum_{i} \frac{\abs{g_{\boldsymbol{k}, q_i, \theta, \lambda, \chi}^{\eta\eta'}}^2 }{\abs{f_{\eta\chi}'(q_i)}} \nonumber \\
    &\cross q_i\left(\coth\frac{\omega_{q_i,\theta}}{2T}-\chi \tanh\frac{E_{\boldsymbol{k},q_i, \theta, \eta}}{2T}\right),
\end{align}
while the real part of the self-energy is obtained using Eq.~\eqref{eq:resigma}.

\section{Self-energy and renormalized Green's function} \label{sec:Results}

\begin{figure*}
    \centering
    \includegraphics[width=\linewidth]{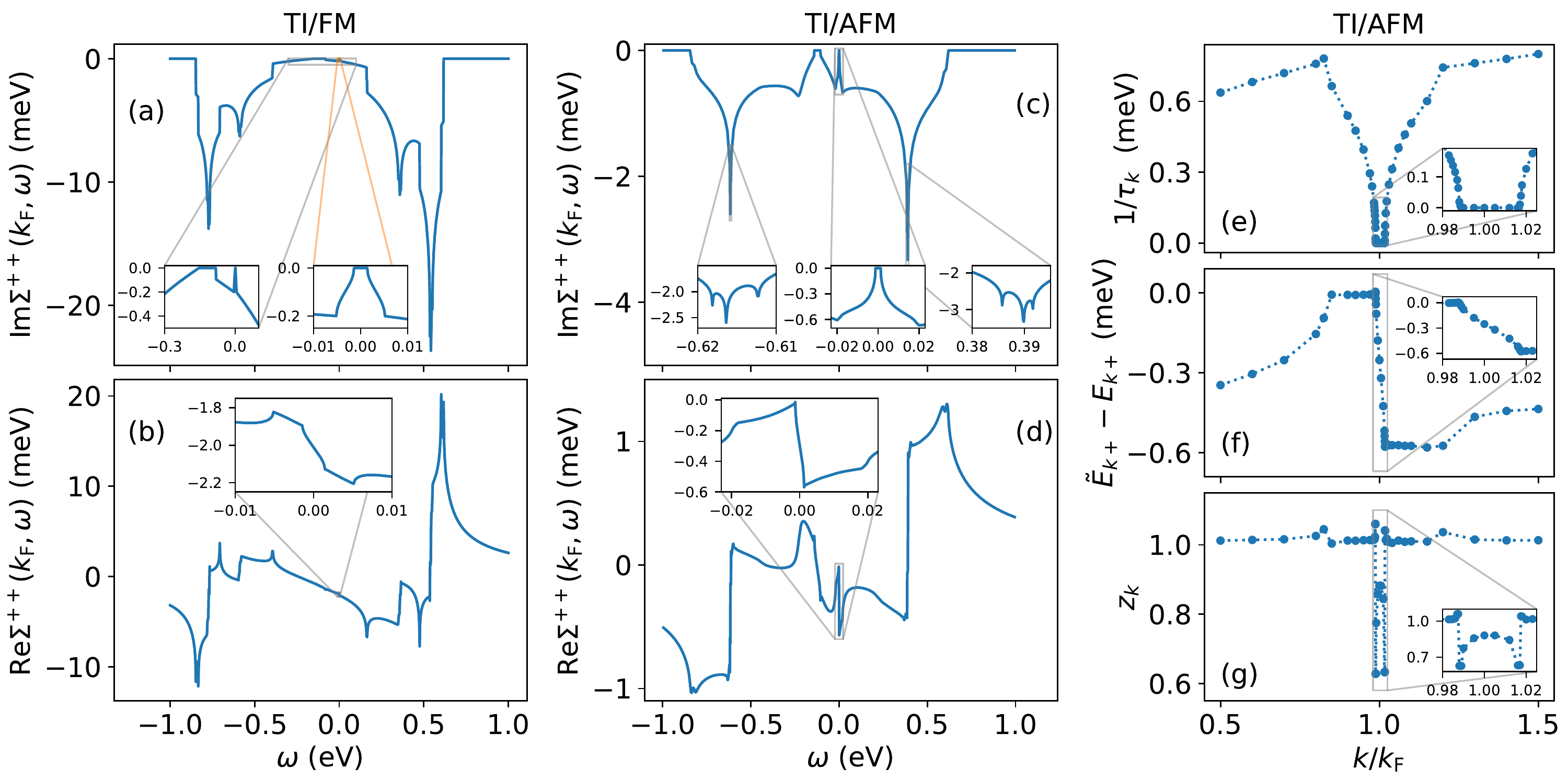}
    \caption{(a) $\Im\Sigma^{++}(\boldsymbol{k}_{\textrm{F}}, \omega)$ and (b) $\Re\Sigma^{++}(\boldsymbol{k}_{\textrm{F}}, \omega)$ for the TI/FM heterostructure, with $k_{\textrm{F}} = \pi/12$, $v_{\textrm{F}} = 429~$meV, $W=0.3v_{\textrm{F}}$, $J = 7~$meV , $\Bar{J} = 18~$meV, $K = J/10$, $S=1$, and $T=10^{-7}~$eV. (c) $\Im\Sigma^{++}(\boldsymbol{k}_{\textrm{F}}, \omega)$ and (d) $\Re\Sigma^{++}(\boldsymbol{k}_{\textrm{F}}, \omega)$ for the TI/AFM heterostructure, with $J_1 = 7~$meV,  $J_2 = 0.05 J_1$, $K = J_1/10^3$, $\Omega = 0$, and otherwise the same parameters. The insets show behaviors that are not easily visible in the main plots. (e) The inverse lifetime, $1/\tau_{\boldsymbol{k}}$, (f) the shift of the excitation spectrum, $\Tilde{E}_{\boldsymbol{k}+}-E_{\boldsymbol{k}+}$, and (g) the quasiparticle residue, $z_{\boldsymbol{k}}$, for the TI/AFM heterostructure with the same parameters. The insets show the behaviors close to the Fermi level. The circles show the calculated points, while the dotted lines are included for visualization. As is mentioned in the text, the results are plotted in the positive $k_x$ direction, with $k_y = 0$. The same applies to the remaining figures plotting results as functions of momentum.}
    \label{fig:sigma}
\end{figure*}

In both the FM case and the AFM case, we assume an electron-doped system with $\mu > 0$. It is then the positive helicity band which crosses the Fermi level, and so we focus on $\Sigma^{++}$ from now on. Due to the lattice nature of our treatment, the system is not isotropic, though both $E_{\boldsymbol{k}\eta}$ and $\omega_{\boldsymbol{q}}$ are nearly isotropic close to the center of the 1BZ. Therefore, our results will be similar in all directions for $\boldsymbol{k}$. The representative direction $k_x \geq 0$, $k_y = 0$, is chosen in all figures, using $k = |\boldsymbol{k}| = k_x$ and $\boldsymbol{k}_{\textrm{F}} = (k_{\textrm{F}}, 0) = (\pi/12, 0)$. With the Fermi momentum fixed, the chemical potential is determined by setting the Fermi energy to zero, $E_{\textrm{F}} \equiv E_{\boldsymbol{k}_{\textrm{F}}, +} = 0$, yielding $\mu = F_{\boldsymbol{k}_{\textrm{F}}}$. The chemical potential should not be too high since the bulk bands will then influence the physics on the TI surface \cite{Hugdal3SCTI}. With the parameters used in this paper, $\mu$ is kept in the region of $100$ to $160~$meV, ensuring that it is reasonable to ignore the bulk bands in the treatment of the TI surface \cite{Hugdal3SCTI}.

The imaginary part of the self-energy is shown as a function of $\omega$ at $\boldsymbol{k} = \boldsymbol{k}_{\textrm{F}}$ in Figs.~\ref{fig:sigma}(a) and \ref{fig:sigma}(c) for the FM case and the AFM case, respectively. The insets show that $|\Im\Sigma^{++}(\boldsymbol{k}_{\textrm{F}}, \omega)|$ is small for small $|\omega|$, i.e.,  close to the Fermi level. This indicates that the fermionic quasiparticles close to the Fermi level are long-lived. The use of $|\Im\Sigma^{++}(\boldsymbol{k}_{\textrm{F}}, \omega)|$ as an indication of the inverse lifetime is made more clear in Sec.~\ref{sec:FLT}. 

We note that for both the FM case and the AFM case there is a drop of $|\Im\Sigma^{++}(\boldsymbol{k}_{\textrm{F}}, \omega)|$ to zero for negative values of $\omega$ comparable to the chemical potential. We find that this extended zero is located around $\omega = -\mu - \omega_{\boldsymbol{q} = \boldsymbol{k}_{\textrm{F}}}$ and that the extent of the zero corresponds to the gap in the excitation spectrum of the TI. For the AFM case, setting $\Omega = 1$ would close the gap, and there would be a single zero at $\omega = -\mu - \omega_{\boldsymbol{q} = \boldsymbol{k}_{\textrm{F}}}$. A similar behavior was found for a Dirac-type fermionic spectrum in Ref.~\cite{elphcselfenergyprlpark} where the self-energy due to EPC is explored in graphene. The suppression of the imaginary part of the self-energy is attributed to the vanishing fermionic density of states (DOS) at the Dirac points. The same explanation holds here, with the adjustment that for a gapped fermionic excitation spectrum there is a range of energies where the fermionic DOS is zero. Another adjustment is that while Ref.~\cite{elphcselfenergyprlpark} studies optical phonons with a fixed frequency, we here study magnons with momentum dependent frequencies. Naively, this should remove the suppression of $|\Im\Sigma^{++}|$, but, as it turns out, the $\delta$ function involved in calculating the self-energy, $\delta(f_{\eta\chi}(q))$, fixes $q$ to certain values in such a way that the suppression remains. To be specific, when $\omega \approx -\mu - \omega_{\boldsymbol{q} = \boldsymbol{k}_{\textrm{F}}}$, satisfying the $\delta$ function requires $\boldsymbol{q} \approx -\boldsymbol{k}_{\textrm{F}}$, fixing the magnon frequencies to $\omega_{\boldsymbol{q}} \approx \omega_{\boldsymbol{q} = \boldsymbol{k}_{\textrm{F}}}$. Hence, scatterings with fermions close to the Dirac point are the relevant processes, just as in Ref.~\cite{elphcselfenergyprlpark}.

Another similarity of the FM and AFM cases is the large peaks in $|\Im\Sigma^{++}(\boldsymbol{k}_{\textrm{F}}, \omega)|$ located at intermediate $|\omega|$. This is attributed to energy ranges around the extrema of the fermionic excitation energies, where $E_{\boldsymbol{k}\eta}$ values are relatively flat, giving a large DOS. Combined with the fact that all magnons are energetically available, this gives a significant increase in the available electron-magnon scattering channels. Also note that these peaks in $|\Im\Sigma^{++}(\boldsymbol{k}_{\textrm{F}}, \omega)|$ are stronger for the FM case than for the AFM case at the same parameters. In the FM case, $\omega_{\boldsymbol{q}}$ values, and to some extent $E_{\boldsymbol{k}\eta}$ values, are more slowly varying, further increasing the available scattering channels.

The real part of the self-energy at $\boldsymbol{k} = \boldsymbol{k}_{\textrm{F}}$ is shown in Figs.~\ref{fig:sigma}(b) and \ref{fig:sigma}(d) for the FM case and the AFM case, respectively. All the exotic behavior found in $\Re\Sigma^{++}(\boldsymbol{k}_{\textrm{F}}, \omega)$ can be traced back to rapid changes of $\Im\Sigma^{++}(\boldsymbol{k}_{\textrm{F}}, \omega)$ at the same values of $\omega$. The real part of the self-energy can be used as an indication of the shift in the fermion spectrum. This is made more clear in Sec.~\ref{sec:FLT}.

\begin{figure*}
    \centering
    \includegraphics[width=\linewidth]{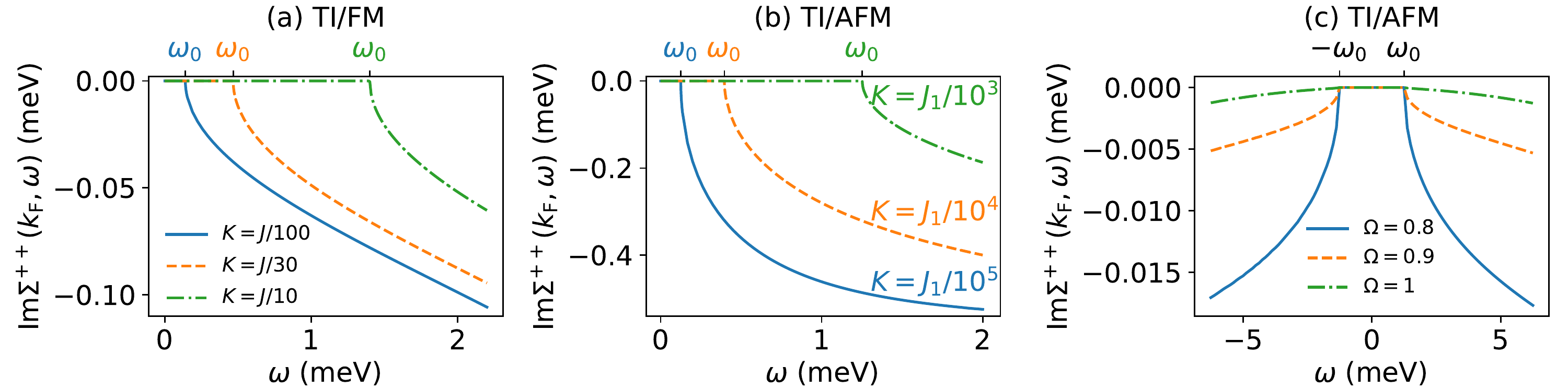}
    \caption{$\Im\Sigma^{++}(\boldsymbol{k}_{\textrm{F}}, \omega)$ for the (a) TI/FM and (b) TI/AFM heterostructure, with the same parameters as those in Fig.~\ref{fig:sigma} except that we vary $K$. One can clearly see that the quick increase in $|\Im\Sigma^{++}(\boldsymbol{k}_{\textrm{F}}, \omega)|$ sets in at $|\omega| = \omega_{\boldsymbol{q} = 0} = \omega_0$, i.e., at the gap in the magnon spectrum. This is shown by the upper ticks, where the color indicates the corresponding curve. (c) $\Im\Sigma^{++}(\boldsymbol{k}_{\textrm{F}}, \omega)$ for the TI/AFM heterostructure with $K = J_1/10^3$, varying $\Omega$, and otherwise the same parameters as those in Fig.~\ref{fig:sigma}. We see that any $\Omega<1$ shows the same behavior as $\Omega = 0$, except that the effect is weaker, in the sense that $|\Im\Sigma^{++}(\boldsymbol{k}_{\textrm{F}}, \omega)|$ is smaller at the same value of $\omega$. }
    \label{fig:fermi}
\end{figure*}

Fig.~\ref{fig:fermi} explores the behavior close to the Fermi level, i.e., for small $|\omega|$, in greater detail. In Fig.~\ref{fig:fermi}(a), we focus on the TI/FM heterostructure and show how the easy-axis anisotropy of the FM, $K$, and in turn the gap in the magnon spectrum, $\omega_{\boldsymbol{q}=0} = 2KS$, determines the extent of $\omega$ values where $|\Im\Sigma^{++}(\boldsymbol{k}_{\textrm{F}}, \omega)|$ is exponentially suppressed. 
The extent of this thermal suppression turns out to be exactly $|\omega| < \omega_{\boldsymbol{q}=0}$, due to the fact that the temperature is kept significantly lower than the gap in the magnon spectrum. The low temperature, $T \ll \omega_{\boldsymbol{q}=0}$, means that very few fermion states with energy between $E_\textrm{F}$ and $E_\textrm{F} + \omega_{\boldsymbol{q}=0}$ are occupied, while almost all states below $E_\textrm{F}$ are occupied. Hence, for fermionic quasiparticles with energy $|\omega| < \omega_{\boldsymbol{q}=0}$ the Pauli principle ensures that there are very few available electron-magnon scattering channels \cite{elphcselfenergyprlpark, elphcselfenergyprltse}.
Once $|\Im\Sigma^{++}(\boldsymbol{k}_{\textrm{F}}, \omega)|$ becomes nonzero for $\omega > \omega_{\boldsymbol{q}=0}$, it increases as $(\omega-\omega_{\boldsymbol{q}=0})^\nu$, where $\nu < 1$. This is non-Fermi liquid behavior, although the extended suppression of $\Im\Sigma^{++}(\boldsymbol{k}_{\textrm{F}}, \omega)$ closer to $\omega = 0$, ensures that the system behaves as a Fermi liquid close to the Fermi level. A similar non-Fermi liquid behavior was found in Ref.~\cite{kargarianprl}, considering an ungapped magnon spectrum. We have shown that introducing an easy-axis anisotropy, and so a gap in the magnon spectrum, can move the non-Fermi liquid behavior away from the Fermi level.

In Fig.~\ref{fig:fermi}(b) the same effect is shown for the TI/AFM heterostructure. 
Since, with $J_1 = J$ and the same $K$, the gap in the AFM magnon spectrum is significantly larger than that in the FM case, a lower degree of easy-axis anisotropy is needed in the AFM case to stabilize the fermionic state close to the Fermi level. 
Also notice that, with comparable gaps, the self-energy increases more rapidly in the AFM case with $\Omega = 0$ than in the FM case. This is due to the increase in EMC for small $\boldsymbol{q}$ with $\Omega \ll 1$ not found in the FM case. Otherwise, the behavior is similar to that of the FM case.

We have thus far considered the strongest possible coupling with $\Omega = 0$ for the AFM case. In Fig.~\ref{fig:fermi}(c) the behavior is now shown for $\Omega > 0$. The general behavior is similar for all $\Omega < 1$, even though $|\Im\Sigma^{++}(\boldsymbol{k}_{\textrm{F}}, \omega)|$ decreases as $\Omega$ is increased. This makes sense, since the EMC decreases when $\Omega$ is increased. Even for $\Omega = 1$ we find an initial fast increase of $|\Im\Sigma^{++}(\boldsymbol{k}_{\textrm{F}}, \omega>\omega_{\boldsymbol{q}=0})|$ in the sense that is goes like $(\omega-\omega_{\boldsymbol{q}=0})^\nu$ with $\nu <1$, but the behavior quickly transitions to a $\nu > 1$ type of increase.

\subsection{Renormalized excitation spectrum, lifetime of quasiparticles, and quasiparticle residue} \label{sec:FLT} 
For the upper helicity band, the renormalized Green's function is given as \cite{BruusFlensberg, migdal}
\begin{equation}\label{eq:renormGreens}
    G^+(\boldsymbol{k}, \omega) = \frac{1}{\omega - E_{\boldsymbol{k}+}-\Sigma^{++}(\boldsymbol{k}, \omega)}.
\end{equation}
Using Fermi liquid theory \cite{BruusFlensberg}, this can be rewritten as
\begin{equation}
    G^+(\boldsymbol{k}, \omega) = \frac{z_{\boldsymbol{k}}}{\omega-\Tilde{E}_{\boldsymbol{k}+}+i/\tau_{\boldsymbol{k}}},
\end{equation}
where the renormalized excitation spectrum $\Tilde{E}_{\boldsymbol{k}+} $ is the solution of
\begin{equation}
\label{eq:renormalized}
    \Tilde{E}_{\boldsymbol{k}+}  = E_{\boldsymbol{k}+} + \Re\Sigma^{++}(\boldsymbol{k}, \Tilde{E}_{\boldsymbol{k}+} ) ,
\end{equation}
the quasiparticle lifetime $\tau_{\boldsymbol{k}}$ is given by
\begin{equation}
\label{eq:tau}
    \frac{1}{\tau_{\boldsymbol{k}}} =  - \frac{\Im\Sigma^{++}(\boldsymbol{k}, \Tilde{E}_{\boldsymbol{k}+} )}{1-\left.\pdv{\Re\Sigma^{++}}{\omega}\right\vert_{\Tilde{E}_{\boldsymbol{k}+} }},
\end{equation}
and the quasiparticle residue $z_{\boldsymbol{k}}$ is
\begin{equation}
\label{eq:residue}
    z_{\boldsymbol{k}} = \frac{1}{1-\left.\pdv{\Re\Sigma^{++}}{\omega}\right\vert_{\Tilde{E}_{\boldsymbol{k}+} }}.
\end{equation}

We choose to calculate these quantities for the TI/AFM heterostructure with an uncompensated interface. The increased EMC, combined with the fact that the easy-axis anisotropy is more effective in producing a gap in the magnon spectrum, is the reason we find the AFM case to be more interesting than the FM case and hence worth exploring in greater detail.

The inverse quasiparticle lifetime is shown in Fig.~\ref{fig:sigma}(e). As indicated by the imaginary part of the self-energy in Fig.~\ref{fig:sigma}(c), the inverse quasiparticle lifetime is exponentially suppressed around the Fermi level, ensuring that the fermionic states are long-lived excitations of the system. Meanwhile, once we move far enough away from the Fermi level, i.e., an energy amount determined by the gap in the magnon spectrum, the inverse lifetime increases rapidly. In other words, the quasiparticle lifetime decreases substantially, and the stability of the fermionic states becomes questionable. 

Note that the extent of the exponential suppression of the inverse lifetime is not symmetric about $k = k_{\textrm{F}}$. This can be understood from the rapid change in the shift of the excitation spectrum, $\Tilde{E}_{\boldsymbol{k}+}-E_{\boldsymbol{k}+}$, for $k$ close to $k_{\textrm{F}}$ shown in Fig.~\ref{fig:sigma}(f). Our calculations predict some sharp ``kinks'' in the renormalized excitation spectrum, which should in principle be observable when measuring the occupied electronic states using ARPES. However, the effects are too small at the chosen parameters to be measurable in current experimental setups \cite{Claessen2004, Claessen2009, Borisenko2012one, Rosenzweig2020overdoping, Iwasawa2020high, Tamai2013spin, Rosenzweig2019tuning}. 
Additionally, we note that the bare band $E_{\boldsymbol{k}+}$ varies over an energy range of the order of $100~$meV for the same momenta, and so the obtained renormalization of the energy can be classified as very weak.

Fig.~\ref{fig:sigma}(g) shows the quasiparticle residue. 
Around the Fermi level we have $z_{\boldsymbol{k}} > 0$, ensuring that the system behaves like a Fermi liquid. The physical interpretation is that a large part of the original fermionic quasiparticle behavior exhibited by the $\psi_{\boldsymbol{k}}$ operators remains after taking the EMC interaction terms in Eq.~\eqref{eq:HintAFM} into account. Meanwhile, further away from the Fermi level $z_{\boldsymbol{k}} > 1$, which is somewhat unusual. It does not seem to make sense that the quasiparticle residue is greater than $1$. The mathematical explanation is that $\Re\Sigma^{++}(\boldsymbol{k}, \omega)$ is an increasing function of $\omega$ around $\omega = \Tilde{E}_{\boldsymbol{k}+} $ at the same values of $\boldsymbol{k}$ where $z_{\boldsymbol{k}} > 1$. Physically, this connects to the non-Fermi liquid behavior exhibited by $|\Im\Sigma^{++}(\boldsymbol{k}, \omega)|$ once it starts increasing rapidly.  The interpretation of $z_{\boldsymbol{k}}$ as a quasiparticle residue is a result of Fermi liquid theory, which may not be valid at the parameters where $z_{\boldsymbol{k}} > 1$.

In Figs.~\ref{fig:sigma}(e)-\ref{fig:sigma}(g), $0.5 \leq k/k_{\textrm{F}} \leq 1.5$, meaning that $-\mu/2 \lesssim  E_{\boldsymbol{k}+} \lesssim \mu/2 \approx 57~$meV. 
We now compare $1/\tau_{\boldsymbol{k}}$ in Fig.~\ref{fig:sigma}(e) to $-\Im\Sigma^{++}(\boldsymbol{k}_{\textrm{F}}, \omega)$ in Fig.~\ref{fig:sigma}(c) and $\Tilde{E}_{\boldsymbol{k}+}-E_{\boldsymbol{k}+}$ in Fig.~\ref{fig:sigma}(f) to $\Re\Sigma^{++}(\boldsymbol{k}_{\textrm{F}}, \omega)$ in Fig.~\ref{fig:sigma}(d) for $-\mu/2 \lesssim  \omega \lesssim \mu/2$. 
Though not exactly the same, it is clear that the plots of the self-energy as functions of $\omega$ at the Fermi momentum provide a good indication of the results for $1/\tau_{\boldsymbol{k}}$ and $\Tilde{E}_{\boldsymbol{k}+}-E_{\boldsymbol{k}+}$ as functions of $k$. Hence, the results for $\Sigma^{++}(\boldsymbol{k}_{\textrm{F}}, \omega)$ in Figs.~\ref{fig:sigma}(a) and \ref{fig:sigma}(b) for the FM case give a good indication of how the quasiparticle lifetime and the renormalized excitation spectrum behave for that system as well. 
For the same values of $k/k_{\textrm{F}}$, the inverse lifetime is smaller and the shift in the excitation spectrum is larger in the FM case. However, the renormalization of the energy is still small compared to the energy range of the bare band.

\section{Towards experimental measurement} \label{sec:ARPES}
We have thus far considered material parameters relevant for the theoretical calculations in Ref.~\cite{EirikTIFMAFM} and shown that the assumption of low renormalization of the fermionic state makes sense, at least at a low temperature of $T = 10^{-7}~$eV, corresponding to $T \approx 10^{-3}~$K. ARPES experiments are, however, typically performed at significantly higher temperatures \cite{Claessen2004, Claessen2009, Borisenko2012one, Rosenzweig2020overdoping, Iwasawa2020high, Tamai2013spin, Rosenzweig2019tuning}. We choose $T = 2.2~$meV, corresponding to $T \approx 25$ K, as a temperature which is readily achievable experimentally. This temperature is much greater or comparable to the magnon gaps we have considered thus far. Hence, more electron-magnon scattering channels become available close to the Fermi level, and the suppression of $|\Im\Sigma^{++}(\boldsymbol{k}_{\textrm{F}}, \omega)|$ close to $\omega = 0$ is lost. In order to increase the magnon gaps, we increase the nearest-neighbor coupling slightly and consider higher degrees of easy-axis anisotropy.

An interfacial exchange coupling of $\Bar{J} = 18~$meV is similar to the values used in several theoretical papers previously \cite{EirikTIFMAFM, thingstad2021eliashberg, AFMNMAFMArnePRB, FMNMFMArnePRB}. These values are, among other measurements, based on an experiment involving a FM deposited on a SC, where the effect of the FM on the superconducting transition is used to estimate the interfacial exchange coupling \cite{ExpWeakJbar}. Alternatively, comparable values have been estimated from measurements of an effective Zeeman field at FM/SC and FM/normal metal (NM) interfaces \cite{FMNMFMArnePRB, ExpWeakJbarZeeman, ExpWeakJbarZeeman2}. To get values of the self-energy that are measurable in ARPES, we consider a system where the interfacial exchange coupling is significantly larger, namely, $\Bar{J} = 100~$meV. Similar values of $\Bar{J}$ are used in Refs.~\cite{magetizationdynamics, strongJbarPRL}, based on an experiment with magnetic impurities in the bulk of a TI. There, the exchange interaction between magnetic impurities and charge carriers is estimated from the behavior of the magnetoresistance \cite{ExpStrongJbar}. Also, a larger value of $\Bar{J}$ is estimated for the FM/NM interface of YIG and gold in Ref.~\cite{FMNMFMArnePRB} based on measurements of the spin-mixing conductance \cite{ExpStrongJbarYIG1, ExpStrongJbarYIG2, ExpFMYIG}. It is emphasized that $\Bar{J} = 100~$meV is not chosen with some specific set of materials in mind, but as a general value that should in principle be achievable for TI/(A)FM interfaces based on the examples listed here. Increasing $\Bar{J}$ simultaneously increases the gap in the fermion spectrum, and so the chemical potential corresponding to $k_{\textrm{F}} = \pi/12$ is now $\mu \approx 152~$meV.

Furthermore, we focus on the TI/AFM structure with $\Omega=0$ since this revealed the strongest EMC. It should be noted that there is a discrepancy between our model and the most realistic experimental realization of a completely uncompensated interface \cite{thingstad2021eliashberg}. With $\Omega = 0$, the lattice on the surface of the TI should match one of the sublattices of the AFM, not the original square lattice. This sublattice is still a square lattice, however, its lattice constant is a factor of $\sqrt{2}$ larger. As elaborated in Ref.~\cite{thingstad2021eliashberg}, this has consequences for the size of the Brillouin zone for the fermions on the TI surface as well. Our model was chosen so that it could describe any value of $\Omega$ satisfying $0\leq\Omega\leq1$, at the cost of this discrepancy in the specific case of $\Omega = 0$. We expect that the results regarding increased EMC at $\Omega = 0$, the effect of the magnon gap on the self-energy close to the Fermi level, and the order of magnitude of the self-energy would be similar also if this detail were to be treated more accurately, or if other lattice configurations were studied.

\begin{figure}
    \centering
    \includegraphics[width=\linewidth]{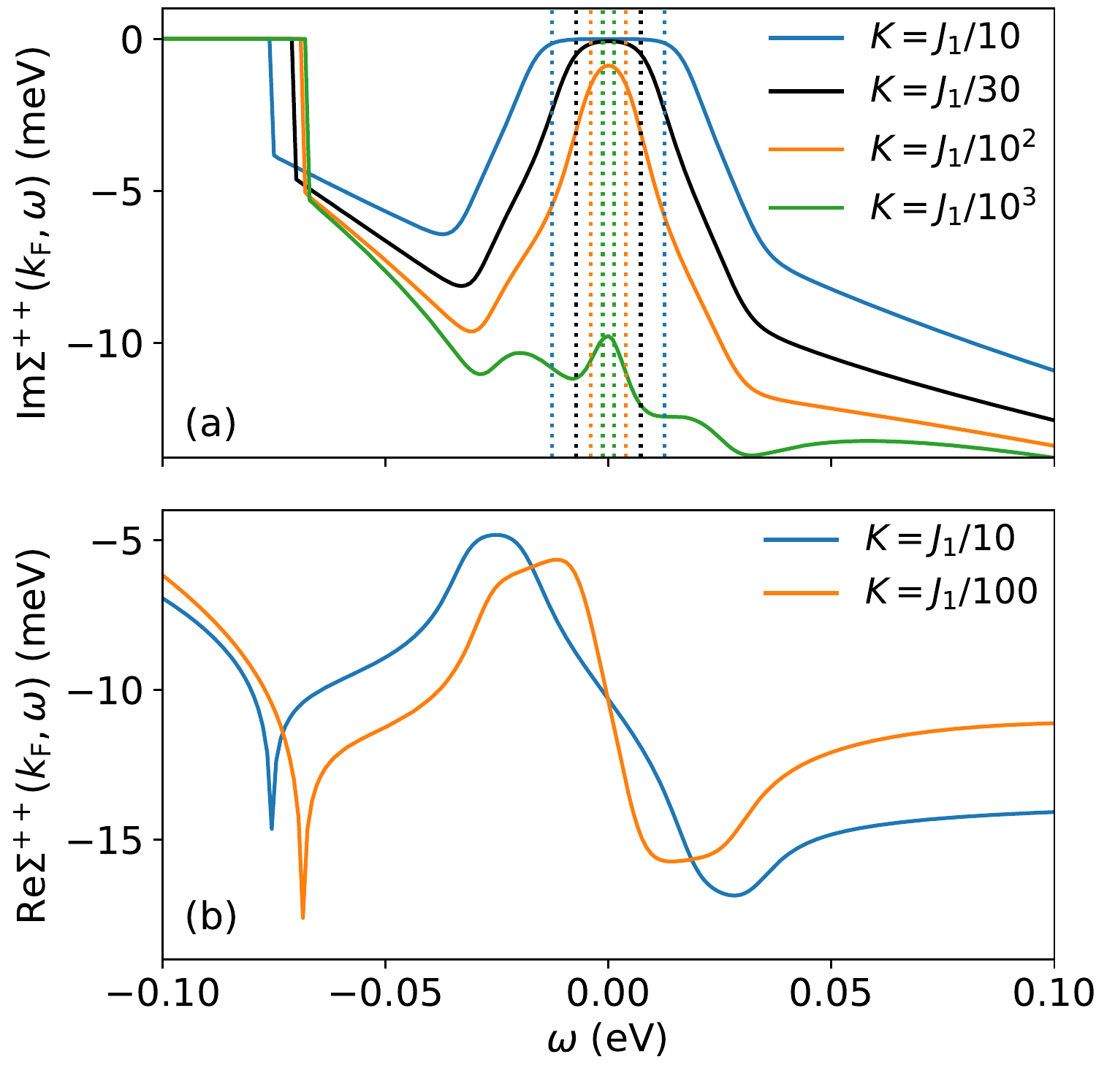}
    \caption{(a) $\Im\Sigma^{++}(\boldsymbol{k}_{\textrm{F}}, \omega)$ for the TI/AFM heterostructure, with $k_{\textrm{F}} = \pi/12$, $v_{\textrm{F}} = 429~$meV, $W=0.3v_{\textrm{F}}$, $J_1 = 10~$meV,  $J_2 = 0.05 J_1$, $\Bar{J} = 100~$meV, $S=1$, $\Omega = 0$, $T=2.2~$meV, and various $K$. The vertical dotted lines show the positions of $\omega = \pm\omega_{\boldsymbol{q}=0}$ for the parameters corresponding to the solid lines of the same color. (b) $\Re\Sigma^{++}(\boldsymbol{k}_{\textrm{F}}, \omega)$ with the same parameters, focusing on two choices for the easy-axis anisotropy. \label{fig:imsAFMT25diffK} \label{fig:resAFMT25} }
\end{figure}

Fig.~\ref{fig:imsAFMT25diffK}(a) shows the imaginary part of the self-energy for the new parameters. Values of $K$ such that the magnon gap is just below and much greater than the temperature have been chosen, showing how the magnon gap affects the self-energy close to the Fermi level. Notice that increasing $K$ decreases $|\Im\Sigma^{++}(\boldsymbol{k}_{\textrm{F}}, \omega)|$ outside the gap region as well. This is because increasing $K$, and so increasing the gap in the magnon spectrum, decreases the maximum magnitude of the Bogoliubov factors $u_{\boldsymbol{q}}$ and $v_{\boldsymbol{q}}$ in Eqs.~\eqref{eq:Bogoliubov_u} and \eqref{eq:Bogoliubov_v}. Hence the EMC is not as strong for larger easy-axis anisotropy.

The corresponding real part of the self-energy is shown in Fig.~\ref{fig:resAFMT25}(b), from now on focusing on two choices for $K$. 
The values of the self-energy shown in these figures should be measurable in ARPES for common energy resolutions at the given temperature \cite{Claessen2004, Claessen2009, Borisenko2012one, Rosenzweig2020overdoping, Iwasawa2020high, Tamai2013spin, Rosenzweig2019tuning}. We also note that the effects of EMC in the TI should dominate over EPC at these parameters, based on the EPC calculations presented in Ref.~\cite{GiraudEgger}. 



From the one-particle Green's function in Eq.~\eqref{eq:renormGreens} one  defines the  \emph{spectral function} \cite{damascelli2004probing}:
\begin{align}\label{eq:spectral}
    &\mathcal{A}(\boldsymbol{k},\omega) = -\pi^{-1}\Im G^+(\boldsymbol{k},\omega) \nonumber \\ &=\frac{-\pi^{-1}\Im\Sigma^{++}(\boldsymbol{k},\omega)}{\bqty{\omega-E_{\boldsymbol{k}+}-\Re\Sigma^{++}(\boldsymbol{k},\omega)}^2+\bqty{\Im\Sigma^{++}(\boldsymbol{k},\omega)}^2}.
\end{align}
In the context of photoexcitation from an interacting $N$-electron system, $\mathcal{A}(\boldsymbol{k},\omega)$ describes the probability of removing an electron with momentum $\boldsymbol{k}$ and energy $\omega$ relative to $E_{\text{F}}$ \cite{hufner2013photoelectron}. The measured intensity of the photoexcitation will be proportional to
\begin{align}\label{eq:ARPESint}
    I(\boldsymbol{k},\omega)\propto &\bqty{\mathcal{M}(\boldsymbol{k},\omega)\cdot\mathcal{A}(\boldsymbol{k},\omega)\cdot g(\boldsymbol{k},\omega)\cdot F_{\text{D}}(\omega)} \nonumber \\ 
    &\ast R_\omega \ast R_k,
\end{align}
where $\mathcal{M}(\boldsymbol{k},\omega)$ describes the photoexcitation matrix elements and $g(\boldsymbol{k},\omega)$ is the electronic DOS. The Fermi-Dirac distribution $F_{\text{D}}(\omega)$ scales the photoemission intensity around the Fermi level. $R_\omega$ and $R_k$ represent the energy and the momentum resolution, respectively.

For nonzero and finite values of $\Sigma^{++}$, $\mathcal{A}(\boldsymbol{k},\omega)$ has a Lorentzian line profile when measured at constant $\boldsymbol{k}$ or $\omega$. $\Im\Sigma^{++}$ is then directly related to the linewidth in an ARPES measurement \cite{gayone2005determining}. Similarly, because the intensity is maximum at
$\omega = \Tilde{E}_{\boldsymbol{k}+}$ [see Eq.~\eqref{eq:renormalized}],
$\Re\Sigma^{++}$ is seen in an ARPES measurement as a renormalization of the occupied band.  Thus, using the calculated Green's function, which includes the bare band and the complex $\Sigma^{++}$, it is possible to simulate an ARPES measurement. In other words carrying out the reverse of the procedure which is typically used to extract an unknown self-energy from measured ARPES data \cite{Pletikosic2012finding,Mazzola2017strong}.

To perform this simulation, we use the self-energy as a function of the renormalized energy band $\Tilde{E}_{\boldsymbol{k}+}$, namely, $\Sigma^{++}(\boldsymbol{k}, \omega = \Tilde{E}_{\boldsymbol{k}+}) = \Sigma^{++}(\Tilde{E}_{\boldsymbol{k}+})$. This is then used as an energy-dependent self-energy whose momentum dependence is neglected, which should be a reasonable approximation \cite{Pletikosic2012finding}. Additionally, a minimal, constant contribution of 5 meV representing electron-impurity scattering has been added to $|\Im\Sigma^{++}|$ to induce a nonzero and more realistic linewidth of the bands \cite{bostwick2007quasiparticle,Bianchi2010electron}.

\begin{figure}
    \centering
    \includegraphics[]{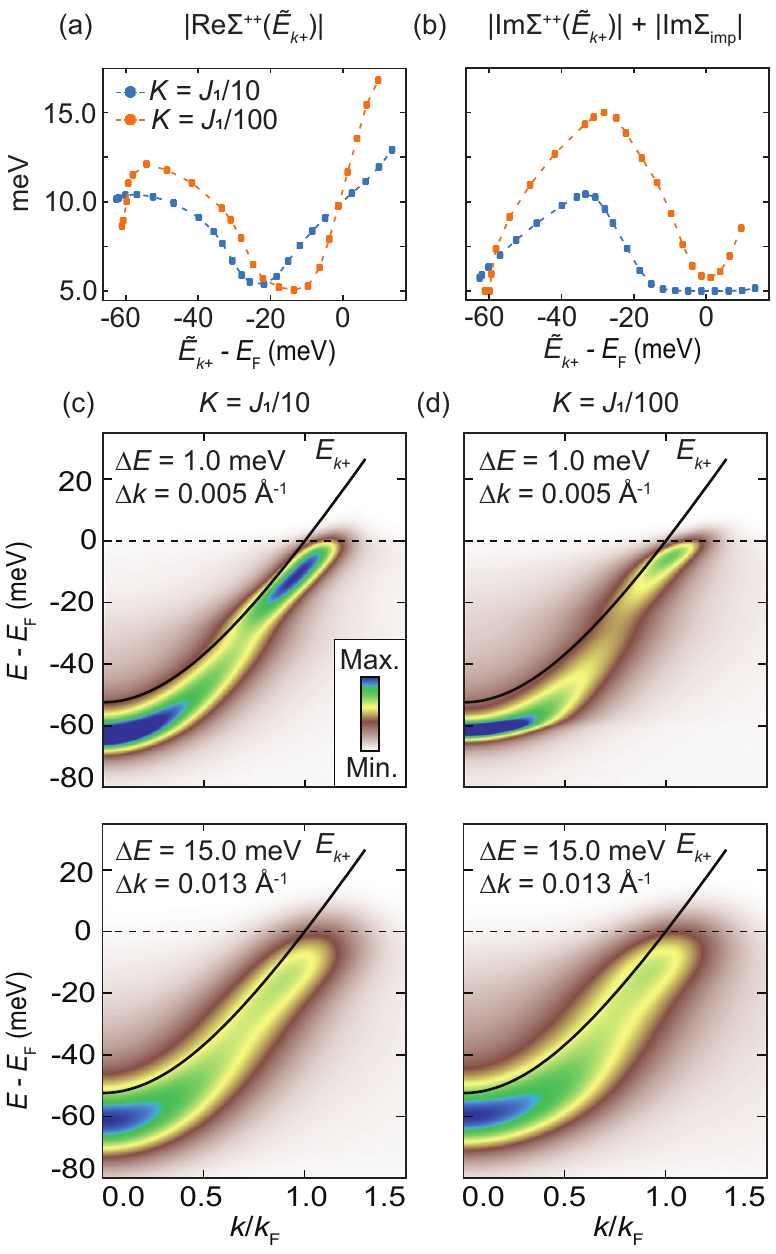}
    \caption{(a,b) Magnitude of the self-energies $\Re\Sigma^{++}(\Tilde{E}_{\boldsymbol{k}+})$ and $\Im\Sigma^{++}(\Tilde{E}_{\boldsymbol{k}+})$ for the TI/AFM heterostructure, with the same parameters as those in Fig.~\ref{fig:imsAFMT25diffK}. A constant energy offset of $5~$meV has been added to the imaginary part to account for a small, but finite broadening due to electron-impurity scattering $\Sigma_{\text{imp}}$. The points show the calculated values, while the dashed lines have been added for visualization. (c,d) Simulations of instrumentally broadened ARPES data showing the positive helicity state $E_{\boldsymbol{k}+}$ renormalized by the self-energies $\Sigma^{++}$ obtained with $K=J_{1}/10$ and $K=J_{1}/100$, respectively. Setting $a = 7~${\AA}, state-of-the-art resolutions for energy, $E$, and momentum, $k$, have been used for the topmost two panels, while the lowermost two panels show the same data with good laboratory-based resolution.}
    \label{fig:ARPESsim}
\end{figure}

$|\Re\Sigma^{++}(\Tilde{E}_{\boldsymbol{k}+})|$ and the shifted $|\Im\Sigma^{++}(\Tilde{E}_{\boldsymbol{k}+})|$ are shown in Figs.~\ref{fig:ARPESsim}(a) and \ref{fig:ARPESsim}(b), respectively, selecting $K=J_{1}/10$ and $K=J_{1}/100$ for the easy-axis anisotropy. Compared to Fermi liquid theory, $|\Im\Sigma^{++}(\Tilde{E}_{\boldsymbol{k}+})|$ should have a functional form similar to $1/\tau_{\boldsymbol{k}}$ as can be seen from Eq.~\eqref{eq:tau}. Meanwhile, $\Re\Sigma^{++}(\Tilde{E}_{\boldsymbol{k}+})$ corresponds to the shift in the excitation spectrum as can be seen in Eq.~\eqref{eq:renormalized}. Notice the significant renormalization of the energy band even for a magnon gap far above the considered temperature. This indicates that a weak-coupling approach to superconductivity in these systems, requiring low renormalization, is best suited at temperatures that are several orders of magnitude smaller than the gap in the magnon spectrum. Otherwise, strong-coupling approaches should be preferred, where the renormalization is taken into account.

Artificially constructed ARPES images showing the positive helicity band $E_{\boldsymbol{k}+}$ of the TI/AFM system with the included self-energy contributions are presented in Figs.~\ref{fig:ARPESsim}(c) and \ref{fig:ARPESsim}(d) for $K=J_{1}/10$ and $K=J_{1}/100$, respectively. 
We emphasize that the plots presented are not measured ARPES data of the system as described, but rather simulated intensity plots based on the theoretically calculated self-energies in Figs.~\ref{fig:ARPESsim}(a) and \ref{fig:ARPESsim}(b).
The simulated plots are produced using Eqs.~\eqref{eq:spectral} and \eqref{eq:ARPESint}, suppressing any variations in the DOS, $g$, and photoexcitation matrix elements $\mathcal{M}$ for simplicity. 
The lattice constant is set to $a = 7~${\AA}, such that our choice of Fermi velocity corresponds to $v_{\textrm{F}} \approx 4.56 \cross 10^{5}~$m/s, which is within the range of reported values \cite{Brune1, Brune2, GiraudEgger, Hugdal3SCTI, Hugdal5SCTI, EirikTIFMAFM}.
Each set of figures is then convolved with two different sets of assumed energy, $E$, and momentum, $k$, resolutions. The upper panels include state-of-the-art, synchrotron ARPES resolutions ($\Delta{E}=1$~meV, $\Delta{k}=0.005$~\AA$^{-1}$) \cite{Borisenko2012one,Rosenzweig2020overdoping,Iwasawa2020high}. The lower panels include good laboratory-based resolutions, (i.e., Specs Phoibos 150 analyzer and non-monochromated He I source; $\Delta E=15$~meV, $\Delta{k} = 0.013$~\AA$^{-1}$) \cite{Tamai2013spin,Rosenzweig2019tuning}.

In the case of excellent instrumental resolutions, both the renormalization and the variation in linewidth because of the magnon interaction are readily observable. For the majority of the energy values the band appears at higher $k$ relative to the undressed dispersion $E_{\boldsymbol{k}+}$, signaling an increased effective mass, $m^{*}$, for electrons in the occupied states. A ``kink'' in the band structure appears around $40-60~$meV below $E_{\text{F}}$, coinciding with the increased $\abs{\Re\Sigma^{++}}$ in this energy range seen from Fig.~\ref{fig:ARPESsim}(a). Furthermore, broadening of the bands is evident from the decreased photoemission intensity between $20$ and $50~$meV below $E_{\text{F}}$, being minimal around $30~$meV below $E_{\text{F}}$ where $\abs{\Im\Sigma^{++}}$ is at its maximum [Fig.~\ref{fig:ARPESsim}(b)].

The same characteristics as described can also be seen from the plots simulated using ``home laboratory'' resolutions. Although being harder to resolve by eye, measures of $\Re\Sigma^{++}$ and $\Im\Sigma^{++}$ would still be straightforward to extract using the analytic approach described in Refs.~\cite{Pletikosic2012finding,Mazzola2017strong}. Thus, we conclude that the predicted self-energy effects due to EMC should be readily observable in a real ARPES experiment using an instrumental setup of reasonable performance.

The results presented here are naturally dependent on the choice of material parameters. For instance, the interfacial exchange coupling $\Bar{J}$, for which a wide range of values has been proposed \cite{EirikTIFMAFM, thingstad2021eliashberg, AFMNMAFMArnePRB, FMNMFMArnePRB, magetizationdynamics, strongJbarPRL}, is directly correlated to the magnitude of the self-energy. Hence, given a TI/FM or TI/AFM heterostructure, we expect that one could compare measured ARPES spectra to the calculations presented here in order to verify the presence and magnitude of the interfacial exchange coupling.
To our knowledge, $\Bar{J}$ has not been measured for either of the heterostructures presented in this paper. If ARPES is performed using the ``home laboratory'' energy and momentum resolutions presented here, $\Bar{J}$ cannot be much lower than 100~meV for this method to succeed. However, a state-of-the-art (synchrotron) ARPES setup should in principle have sufficient resolutions to measure lower values of $\Bar{J}$. Note also that the spin quantum number will affect the magnitude of the self-energy, and the value of $S$ in an experimental realization could be different from $S=1$, as chosen in the figures.

\section{Conclusion} \label{sec:Con}
We have explored self-energy effects on the surface of a topological insulator due to magnetic fluctuations in an adjacent ferromagnet or antiferromagnet. Useful applications of such systems include superconductivity and magnetoelectric effects. In such cases it is often important that the fermionic quasiparticles on the surface of the topological insulator are long-lived excitations. In weak-coupling approaches to superconductivity it is also required that the renormalization of the excitation energies is weak. We have shown how an easy-axis anisotropy in the magnetic insulators can be used to increase the lifetime of the quasiparticles close to the Fermi level. Additionally, we reported a set of parameters where the assumption of weak renormalization of the energy is valid. 
Finally, we studied a system at higher temperature and with a stronger interfacial exchange coupling, giving self-energies measurable in ARPES. We suggest that these calculations could be used, upon comparison to experimental results, e.g., to measure the magnitude of the interfacial exchange coupling. 
Additionally, we find that a greater easy-axis anisotropy is needed at higher temperatures to increase the lifetime of the quasiparticles on the surface of the topological insulator. However, the energy renormalization remains significant. 
Therefore, strong-coupling approaches to superconductivity will in general be needed in these systems, unless one is interested solely in low-temperature results.  

\section*{Acknowledgments}
We acknowledge funding from 
the Research Council of Norway Project No. 250985, ``Fundamentals of Low-Dissipative Topological Matter", and  
the Research Council of Norway through its Centres of Excellence funding scheme, Project No. 262633, ``QuSpin."

\appendix

\section{Tadpole diagram} \label{sec:tadpole}

\begin{figure}[htb]
    \centering
    \includegraphics[width=0.7\linewidth]{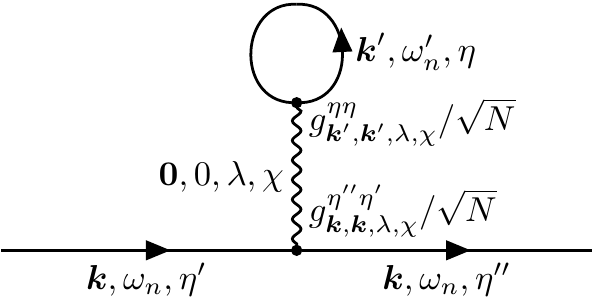}
    \caption{The tadpole diagram is also relevant for second-order EMC. Momentum and energy conservation fixes $\boldsymbol{q} = \boldsymbol{0}$ and $\omega_\nu = 0$ for the magnon. Unlike EPC \cite{GiraudEgger}, the coupling constants remain nonzero. However, it turns out that $g_{\boldsymbol{k}',\boldsymbol{k}',\lambda,\chi}^{\eta\eta}$ is antisymmetric under inversion of $\boldsymbol{k}'$, explaining why this Feynman diagram gives zero contribution to the self-energy.}
    \label{fig:tadpole}
\end{figure}

In this appendix, we show that the tadpole diagram shown in Fig.~\ref{fig:tadpole} gives zero contribution to the self-energy. Denoting this contribution $\Sigma_{\textrm{T}}^{\eta''\eta'}$, we have
\begin{align}
    &\Sigma_{\textrm{T}}^{\eta''\eta'}(\boldsymbol{k}) = \frac{2}{N}\sum_{\boldsymbol{k}'}T\sum_{\omega'_n}\sum_{\lambda, \chi, \eta = \pm} D_0^\chi(\boldsymbol{0}, 0) G_0^\eta(\boldsymbol{k}', \omega'_n) \nonumber \\
    & \mbox{\qquad\qquad\qquad\qquad\qquad\qquad\qquad} \cross g_{\boldsymbol{k}, \boldsymbol{k}, \lambda, \chi}^{\eta''\eta'}g_{\boldsymbol{k}',\boldsymbol{k}',\lambda,\chi}^{\eta\eta} \nonumber \\
    &=  \frac{-2}{N}\sum_{\boldsymbol{k}'}\sum_{\lambda, \chi, \eta = \pm} \frac{F_{\textrm{D}}(E_{\boldsymbol{k}'\eta})}{\omega_{\boldsymbol{q}=0}}g_{\boldsymbol{k}, \boldsymbol{k}, \lambda, \chi}^{\eta''\eta'}g_{\boldsymbol{k}',\boldsymbol{k}',\lambda,\chi}^{\eta\eta}.
\end{align}
Notice that $E_{\boldsymbol{k}'\eta}$ is inversion symmetric in $\boldsymbol{k}'$.
Moreover, by inspecting Eqs.~\eqref{eq:Qcoeff}, \eqref{eq:Qcoeff2}, \eqref{eq:HintFM}, and \eqref{eq:HintAFM}, it becomes clear that $g_{\boldsymbol{k}',\boldsymbol{k}',\lambda,\chi}^{\eta\eta}$ will, for both the TI/FM heterostructure and the TI/AFM heterostructure, always contain a combination $(F_{\boldsymbol{k}'}+B_{\boldsymbol{k}'})(C_{\boldsymbol{k}'}\pm iD_{\boldsymbol{k}'})/N_{\boldsymbol{k}'}$ which is antisymmetric under inversion of $\boldsymbol{k}'$. Therefore, the summand in $\Sigma_{\textrm{T}}^{\eta''\eta'}$ is inversion antisymmetric, yielding $\Sigma_{\textrm{T}}^{\eta''\eta'} = 0$.

\section{Details of \texorpdfstring{$\delta$}{delta}-function treatment} \label{sec:delta} 
There are three cases that require some care when treating the $\delta$ function in the imaginary part of the self-energy, namely, if $q = 0$ is a root of $f_{\eta\chi}(q)$, if $q = c(\theta)$ is a root,  or if double roots appear.
With $K > 0$, we find that any roots at $q=0$ give zero contribution. Only half the borders of the 1BZ/RBZ are included in the sum over $\boldsymbol{q}$. If there is a zero there, the procedure is exactly the same as for a zero at $0 < q < c(\theta)$ except for a factor $1/2$, since the zero is at the edge of the integration interval. For notational convenience, this detail is left out of Eqs.~\eqref{eq:numericimsigma} and \eqref{eq:numericimsigmaAFM}.

Double roots appear if $f_{\eta\chi}'(q) = 0$, and the method we have presented fails. The occurrence of double roots generally happens at a finite set of distinct values of $\theta$. As we approach a double root by varying $\theta$, two distinct roots move closer to each other. Hence, the derivatives $f_{\eta\chi}'(q)$ at these roots approach zero and the integrand in the $\theta$ integral diverges.  
Numerically, we split up the integration interval for $\theta$ to handle such improper integrals. 

\bibliography{main.bbl}

\end{document}